\documentclass[12pt,preprint]{aastex}
\usepackage{amsmath}

\slugcomment{ApJ, in press}

\begin{document}
\title{The Red Rectangle: Its Shaping Mechanism and its Source of Ultraviolet Photons}
\author{Adolf N. Witt\altaffilmark{1}, Uma P. Vijh\altaffilmark{1}, L. M. Hobbs\altaffilmark{2},  Jason P. Aufdenberg\altaffilmark{3} , Julie A. Thorburn\altaffilmark{2}, \and Donald G. York\altaffilmark{4}}
\altaffiltext{1}{Ritter Astrophysical Research Center, University of Toledo, Toledo, OH 43606}
\altaffiltext{2}{Yerkes Observatory, The University of Chicago, Williams Bay, WI 53191}
\altaffiltext{3}{Physical Sciences Department, Embry-Riddle Aeronautical University, Daytona Beach, FL 32114}
\altaffiltext{4}{Department of Astronomy \& Astrophysics, University of Chicago, Chicago, IL 60637}

\begin{abstract}
The proto-planetary Red Rectangle nebula is powered by HD~44179, a spectroscopic binary (P = 318 d), in which a luminous post-AGB component is the primary source of both luminosity and current mass loss.  Here, we present the results of a seven-year, eight-orbit spectroscopic monitoring program of HD~44179, designed to uncover new information about the source of the Lyman/far-ultraviolet continuum in the system as well as the driving mechanism for the bipolar outflow producing the current nebula.  Our observations of the H$\alpha$ line profile around the orbital phase of superior conjunction reveal the secondary component to be the origin of the fast (max. v $\sim560$ km s$^{-1}$) bipolar outflow in the Red Rectangle. The outflow was previously inferred on the basis of a single, broad H$\alpha$ emission line profile. The variation of total H$\alpha$ flux from the central H II region with orbital phase also identifies the secondary or its surroundings as the source of the far-ultraviolet ionizing radiation in the system. The estimated mass of the secondary ($\sim0.94$ M$\sun$) and the speed of the outflow suggest that this component is a main sequence star and not a white dwarf, as previously suggested. We identify the source of the Lyman/far-ultraviolet continuum in the system as the hot, inner region (T$_{max} \ge 17,000$ K) of an accretion disk surrounding the secondary, fed by Roche lobe overflow from the post-AGB primary at a rate of about $2 - 5\times10^{-5}$ M$\sun$ yr$^{-1}$. The speed of the accretion-driven, bipolar outflow was found to be strongly modulated by the inter-component separation in the highly eccentric ($e = 0.37$) orbit, with maximum speeds occurring near periastron and minimum speeds found around apastron. The total luminosity of the accretion disk around the secondary is estimated to be at least 300~L$\sun$, about 5\% of the luminosity of the entire system.
\end{abstract}

\keywords{ISM: jets and outflows, stars: AGB and post-AGB, stars: individual (HD~44179), ISM: individual (Red  Rectangle), Physical Data and Processes: accretion , accretion disks}

\section{Introduction}
The Red Rectangle (RR) and its associated star HD~44179 are an extensively studied protoplanetary nebula system that is frequently discussed as a prototype for intermediate-mass stars traversing the brief evolutionary phase between the top of the asymptotic giant branch (AGB) and the onset of the planetary nebula stage, when the star's hot core begins to ionize the expelled circumstellar matter (van Winckel 2003, 2007; Siodmiak et al. 2008; Szczerba et al. 2007). The RR was discovered on the basis of its strong near- and mid-infrared excess (Price \& Walker 1976) and it was subsequently identified with a compact bipolar nebula associated with the post-AGB star HD~44179,  earning  its name on the basis of its rectangular appearance on existing red photographic plates (Cohen et al. 1975). Subsequent high-resolution imaging at optical (Osterbart et al. 1997; Cohen et al. 2004; Vijh et al. 2006) and near-infrared wavelengths (Tuthill et al. 2002) as well as spectroscopic mapping (Schmidt \& Witt 1991) revealed the RR as an X-shaped biconical nebula, emerging from an optically thick disk seen nearly edge-on. We will refer to this disk, resolved by HST, as the circumbinary disk, to distinguish it from the much smaller accretion disk around the photometric secondary in HD~44179, to be discussed later. The circumbinary disk was explored in detail by Men'shchikov et al. (2002); it is estimated to have inner and outer radii of 14 AU and 43,000 AU, respectively, with a thickness of 90 AU. The distance to the RR is uncertain, but recent detailed modeling of the stellar and nebular emission by Men'shchikov et al. (2002) places the object at a distance of 710 pc, which we adopt for the purpose of the present study.

Soon after the discovery of the RR and other similar objects, two different models emerged for the explanation of the bipolar morphology of such nebulae. Icke (1981, 2003) proposed a biconical outflow pattern above a luminous disk as the shaping mechanism, with radiation pressure as the driving force. On the other hand, Morris (1981, 1987) envisioned binary star systems at the centers of bipolar nebulae, with bipolar jets launched by an accretion disk surrounding the secondary as both the driving and shaping force for the nebula. We note that in the case of HD~44179 its nature as a spectroscopic binary was not established until 1995 by van Winckel et al. (1995). In recent years, the binary star model as the cause of bipolar nebulae was further developed by Mastrodemos \& Morris (1998, 1999) and, independently, by Soker (2000, 2005). Two recent papers by Icke (2003) and Soker (2005) continue the long-standing debate about the shaping mechanism for the special case of the RR. Nordhaus \& Blackman (2006) discuss three possible modes of the evolutionary outcome of the common-envelope phase of a close binary including an AGB star (Livio et al. 1979), one of which may describe an earlier stage of evolution of the RR (e.g. Akashi \& Soker 2008), during which the outer circumbinary disk was formed.

Another controversy concerns the nature of the secondary star in the HD~44179 system. Following the initial identification of HD~44179 as a spectroscopic binary by van Winckel et al. (1995),  Waelkens et al. (1996) confirmed the orbital period of $318\pm3$ days and reported the photometric variability of the system with the same period. The phase dependence of the brightness variation was attributed to the variable effective scattering angle under which the primary component of HD~44179 is being viewed by a process of scattering off dust in the upper layers of the optically thick circumbinary disk. This viewing geometry was subsequently confirmed by high-resolution imaging at optical (Osterbart et al. 1997; Cohen et al. 2004) and infrared wavelengths (Tuthill et al. 2002). An important consequence of this indirect view of HD~44179 is that its ``effective inclination'' is considerably smaller than otherwise implied by the near-edge-on orientation of the system, based on the appearance of the circumbinary disk. Waelkens et al. (1996) estimated an ``effective inclination'' of 35\degr\  for HD~44179. When coupled with the mass function of the single-line spectroscopic binary,

\begin{equation}
f(m_1,m_2)=\frac{m_2^3\ sin^3\ i}{(m_1 +m_2)^2}=\frac{(a_1\ sin\ i)^3}{P^2}=0.049 M\sun ,
\end{equation}

\noindent where $P$ is the period  of the binary, $i$ is the inclination of the orbit, $m_1$ and $m_2$ are the masses of the primary and secondary stars, respectively, and $a_1$ is the semi-major axis of the observed primary,  
the mass of the secondary is larger than the current mass of the photometrically dominant post-AGB primary of the system. Waelkens et al. (1996) suggested, therefore, that the secondary is most likely a low-luminosity main sequence star of less than 1 M$\sun$. However, in such a configuration neither the photometrically dominant post-AGB star (T$_{\mathrm{eff}} < 8000$ K) nor the even cooler main-sequence companion star can explain the radio detection of a small, compact HII region at the center of the RR by Jura et al. (1997). Consequently, Men'shchikov et al. (2002) developed a detailed model for the RR system assuming that the secondary is a hot He white dwarf star, which could supply the required Lyman continuum flux for the observed photo-ionization of hydrogen (Jura et al. 1997) and of helium (Kelly \& Latter 1995), as well as supply the far-UV continuum required for the excitation of the extended red emission (ERE; Witt et al. 2006) and the high-excitation CO emission (Sitko 1983; Sitko et al. 2008) observed in the RR. This scenario, however, implies a greatly more complicated evolutionary history of mass loss and exchange, and it is based upon an ``effective inclination'' of 79\degr\ (Men'shchikov et al. 2002), inconsistent with the results of Waelkens et al. (1996).

In this paper we present the results of a program of high-resolution spectroscopic observations, which provide complete phase coverage over eight orbits of the HD~44179 binary from 2001 to 2008. These observations provide the first direct confirmation of the Morris (1987) model for the shaping of the  RR through a jet emanating from the vicinity of the secondary star. The observed jet speed is consistent with the expected outflow from an accretion disk centered on a low-mass main sequence star, as proposed by Waelkens et al. (1996). If such an accretion disk is fed by the mass loss from the post-AGB component of HD~44179 through Roche lobe overflow, the resulting disk temperatures and disk luminosity can account fully for the Lyman continuum luminosity required for the production of the compact HII region at the center of the RR, observed by Jura et al. (1997).

We are reporting the details of our observations in \S2 of this paper. In \S3 we present the evidence supporting our claim that the source of the bipolar outflow as well as the source of the Lyman contininuum are located in the vicinity of the secondary, as it moves through its orbit. In \S4 we discuss the mass of the secondary as well as the process of mass accretion from the post-AGB component. In \S5 we examine the predictions of steady mass-accretion disk models, including the expected disk spectra and luminosities as a function of the mass transfer rate. This is followed by a discussion and conclusions in \S6.

\section{Observations}
\subsection{Details of Observations}
The spectra of HD~44179 for this program were taken with the ARCES echelle spectrograph mounted on the 3.5-m telescope at Apache Point Observatory in New Mexico (Wang et al. 2003). The spectrograph yields spectra covering the spectral range from 3700~\AA\ to 10,000~\AA\ at a resolving power of R = 38,000. A total of 17 spectra were obtained during a time spanning the interval from February, 2001, to February, 2008. The first seven spectra have been previously analyzed for a study of atomic and molecular emission lines in HD~44179 by Hobbs et al. (2004). The instrumental setup and the reduction techniques employed were described previously by Thorburn et al. (2003) and by Hobbs et al. (2004).

\subsection{Reductions and Stellar Radial Velocities}
A summary of the observations is provided in Table~\ref{t-obs}. Column 8 of Table~\ref{t-obs} lists the orbital phase, which has been calculated with respect to an arbitrarily adopted phase zero at JD 2448300, coinciding with the minimum in the RV curve of Waelkens et al. (1996), and a period of 318 days, derived by Waelkens et al. (1996). The last column reports the measured stellar RV values after applying the heliocentric velocity correction. The stellar RV values are based on measurements of three absorption lines with accurately known laboratory wavelengths, Fe II 4923.927~\AA, C I 4932.049~\AA, and C I 5052.167~\AA, and they have random errors of typically $\pm$0.4~km~s$^{-1}$. For the first seven nights indicated in Table 1, the radial velocities presented in column 10 for the K I emission and in column 11 for the AGB star differ by small amounts from the corresponding values given previously by Hobbs et al (2004). In the cases of both the K I and the stellar velocities, the magnitude of only one such correction exceeds 0.5~km\ $\mathrm{s}^{-1},$ or
6\% of a resolution element. The new results supersede the earlier ones and reflect several improvements. The principal change is that the zero-points of the wavelength scales in the reduced spectra have now been derived from telluric lines instead of from the Th/Ar lines of the comparison lamp. In addition, the velocity derived from the [O I] 6300~\AA\ line, which is affected by interfering telluric lines, has no longer been averaged at half weight with the value derived from the K I line, which is more accurate as well as very similar.

The recent literature contains two conflicting values for the orbital period of HD 44179, 318 days (Waelkens et al. 1996) and 322 days (Men'shchikov et al. 2002). While our stellar RV data span more than 8 orbital periods and were obtained more than 10 years after those of Waelkens et al. (1996), we see no noticeable shift in phase either internally or with respect to the Waelkens et al. (1996) RV curve, as long as we adhere to the period of 318 days derived by Waelkens et al. (1996). We, therefore, confirm the original period found by Waelkens et al. (1996) and constrain the uncertainty to $\pm$1 day. Our data do not support the period of 322 days reported by Men'shchikov et al. (2002). This is of significance, because our later interpretation of the spectra of HD~44179 will depend on an accurate assignment of the correct orbital phase to individual spectra. The specific phase values corresponding to the times of periastron, apastron, and inferior and superior conjunction were adopted from the analyses of Waelkens et al. (1996) and Men'shchikov et al. (2002).

\subsection{The H$\alpha$ Profile}
This investigation was motivated primarily by our interest in finding the source of the Lyman continuum and the cause of the bipolar outflow in the RR. Therefore, our analysis has focussed on the study of the H$\alpha$ profile as a function of orbital phase. An earlier study of the H$\alpha$ emission profile in HD~44179, observed on 1994, January 3, by Jura et al. (1997), had revealed a resolved, narrow, symmetric emission spike of FWHM of 20~km~s$^{-1}$ sitting on top of a broad emission plateau extending to $\pm 100$~km~s$^{-1}$ with respect to the central emission spike.  Jura et al. (1997) found the central emission spike wavelength to be stationary, not following the orbital motion of HD~44179, which was confirmed by Hobbs et al. (2004), who detected many optical emission lines from the same narrow-line emitting region.  Jura et al. (1997) attributed the narrow H$\alpha$ spike to the small, compact HII region in the RR, most likely caused by photo-ionization, in light of the small H$\alpha$ line width, while the broad plateau emission was attributed to material of unspecified origin involved in a rapid bipolar outflow. We note that the viewing geometry for HD~44179, as suggested by Waelkens et al. (1996), allows us to see the outflow from top and bottom simultaneously under a viewing angle of about 35\degr with respect to the outflow direction.

\begin{figure}
\centering{
\includegraphics[width=3.5in]{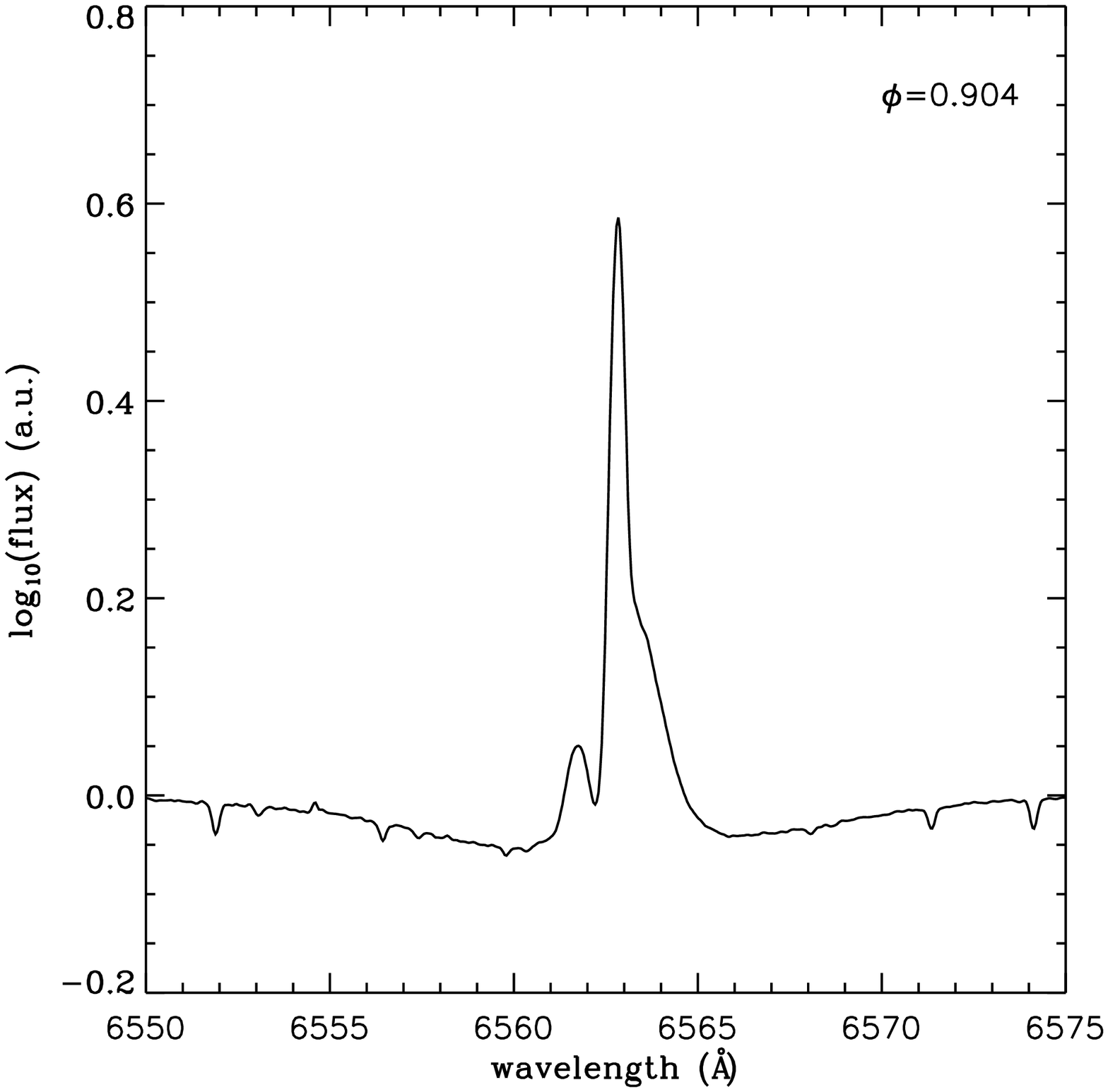}\includegraphics[width=3.5in]{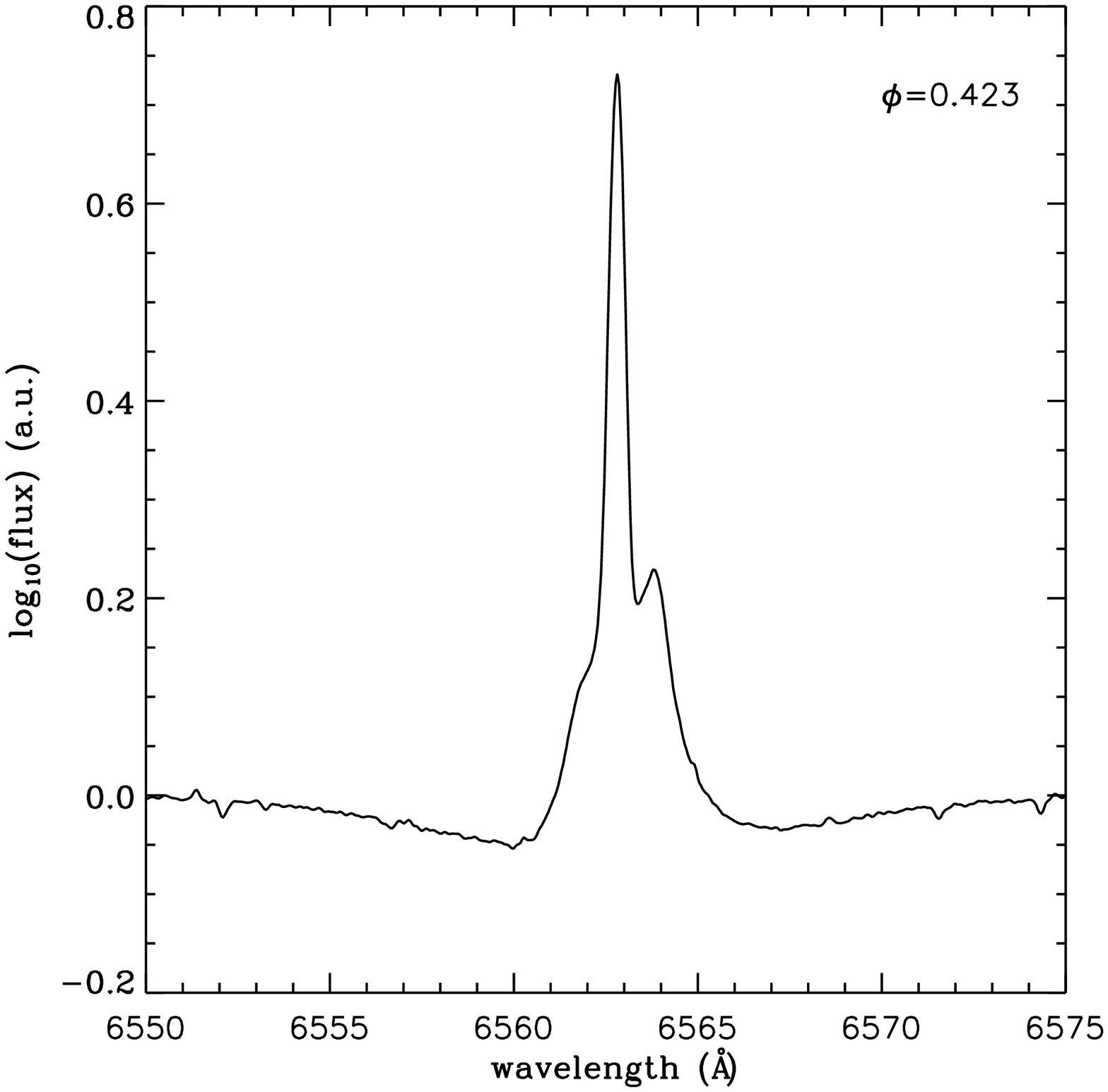}
\caption{(a) Observed H$\alpha$ emission line profile of HD~44179 at phase $\phi$= 0.904, closest to maximum blue-shift of the stellar absorption line spectrum relative to the near-constant H$\alpha$ core emission. (b) Observed H$\alpha$ emission line profile of HD~44179 at phase $\phi$= 0.423, closest to maximum red-shift of the stellar absorption line spectrum relative to the near-constant H$\alpha$ core emission. The observed flux expressed in arbitrary units (a.u.) has been normalized to the distant continuum equal to unity. Note that the flux scale is logarithmic. The faint absorption lines seen in the spectra are telluric in origin.\label{ha7-15}}
}
\end{figure}

The published H$\alpha$ profile of Jura et al. (1997) corresponds to an orbital phase of 0.32, if the reference phase and period of Waelkens et al. (1996) are used. This phase corresponds to a point in the orbit at which the stellar RV has reached about half of its maximum redward excursion relative to the system velocity. In Figure~\ref{ha7-15} we show two H$\alpha$ profiles for orbital phases 0.904 (\#7) and 0.423 (\#15) from our data set, which, respectively, correspond to the phases of maximum blue-shift and maximum red-shift of the stellar (primary) spectrum with respect to the system velocity, in which the H$\alpha$ spike is at rest. Note that the vertical flux scale is logarithmic in order to de-emphasize the great height of the central H$\alpha$ spike and focus attention on the details in the lower part of the H$\alpha$ profiles. The flux scale is normalized to unity in the continuum far from the position of the H$\alpha$ line. In addition to the narrow central emission spike and the broad plateau reported by Jura et al. (1997), our spectra also display a narrow absorption feature, which is either blue-shifted (\#7) or red-shifted (\#15) with respect to the central spike. At intermediate phases with RV values between the two extremes shown here the absorption feature moves across the emission profile, and it becomes undetectable when the stellar RV is near the system velocity. We interpret this feature as the result of the superposition of the stellar H$\alpha$ absorption profile of HD~44179 with a deep, narrow line core on the broad H$\alpha$ emission produced in the out-flowing gas. The changing superposition of the stellar H$\alpha$ absorption line core on the H$\alpha$ emission profile affects the apparent wavelength of the narrow H$\alpha$ spike, making the latter an unreliable indicator of the motion of the gas associated with the central HII region in the RR. 

\subsection{Stellar and Gas Motions}
Hobbs  et al. (2004) identified numerous atomic forbidden and permitted emission lines in the spectrum of HD~44179, which have rather similar FWHM values as the H$\alpha$ spike and which have been attributed to the same emission region. Since HD~44179 is extremely metal-poor ([Fe/H] = - 3.1; van Winckel et al. 1995), atomic emission lines of heavily depleted metals do not suffer from the problem described above (\S2.3) for the H$\alpha$ line. Stellar absorption lines of these metals are correspondingly weak in the spectrum of HD~44179. Therefore, we measured the radial velocity of the unblended KI 7699~\AA\ emission line as a function of phase as an indicator of the motion of the ionized gas in the RR. We show the measured RV values of the luminous primary in HD~44179 and the gas velocities determined from the KI emission line as a function of phase in Figure 2. The numerical values of the KI radial velocity are listed in column 10 of Table 1. It appears that the emitting gas that is producing narrow emission lines is essentially at rest, moving at the system velocity of about (19.2$\pm$0.4)~km~s$^{-1}$ and is not following the orbital motion of the primary. In fact, at a marginally significant level, the gas appears to move in anti-phase with respect to the primary with an amplitude of about 0.6~km~s$^{-1}$.

\section{The Source of the Bipolar Outflow}
\subsection{Analysis of the H$\alpha$ Profile}
\subsubsection{Computation of the Stellar H$\alpha$ Absorption Line Profile}
We employed the PHOENIX code, Version 15.04.00E (Hauschildt 1992, 1993; Hauschildt \& Baron 1995;  Allard \& Hauschildt 1995; Baron et al. 1996; Hauschildt et al. 1996, 1997; Baron \& Hauschildt 1998; Allard et al. 2001; Hauschildt et al. 2001), a very general NLTE stellar atmosphere code, to compute the H$\alpha$ absorption line profile of the HD~44179 post-AGB atmosphere. The model parameters were determined by producing the closest fit to the observed H$\gamma$ profile in HD~44179, the first Balmer line without any significant emission component in the star's spectrum. The parameters adopted where T$_{\mathrm{eff}} = 7700$~K, log g = 1.10, [Z/H] = -3.5, [CNO/H] = -0.5 and a stellar mass of 0.8 M$\sun$, suggested by Waelkens (1996). The H$\alpha$ profile was computed with the parameter set that reproduced the observed H$\gamma$ profile, and it is shown in Figure 3. We scaled the relative intensity of the continuum as a function of phase in accordance with the photometric variability reported by Waelkens (1996), and then,  using the results shown in Figure 2, this H$\alpha$ profile was shifted in wavelength relative to the H$\alpha$ emission spike and subtracted from the observed spectra. The expected result is a pure emission profile which no longer exhibits the narrow absorption reversals due to the stellar absorption line cores visible in the observed spectra (Figure~\ref{ha7-15}). In reality, most of the absorption-corrected emission line profiles still exhibit traces of the absorption line effect, suggesting that the line profile shown in Figure 3 should be somewhat deeper than predicted by the model at the wavelength where the residual absorption dip appears. 

We illustrate a worst case for the phase of maximum stellar blue-shift in Figure 4. The residual blue-shifted absorption dip appears at a radial velocity shift of 28.8~km~s$^{-1}$ relative to the narrow H$\alpha$ core, whereas the relative shift of the stellar spectrum with respect to the system velocity represented by the K I emission is only 10~km~s$^{-1}$ (Table 1). This implies that the problem arises not from an insufficient depth of the line core in the model profile itself but rather from an over-prediction of the intensity about 0.4~\AA\ from the line center; i.e. the actual line core in HD~44179 is slightly wider than that shown in Figure 3. The emission wings in the corrected line profile in Figure 4 can be traced readily out to $\pm$ 10~\AA\ from the line center, corresponding to inclination-corrected maximum outflow speeds near 560~km~s$^{-1}$. 

In light of the facts that the post-AGB component of HD~44179 may not be spherically symmetrical and some fraction of the light of the system may originate in an accretion disk surrounding the secondary, we did not attempt to refine the atmosphere model further.

\subsubsection{H$\alpha$ Profiles at Superior and Inferior Conjunction}
The disk geometry of the RR allows us to view the central binary along two simultaneous lines of sight under angles of approximately 55\degr\ with respect to the orbital plane (Waelkens et al. 1996). The superior and inferior conjunctions, during which the photometric primary is either on the far side or the near side of its orbit, respectively, occur during the phases of 0.59 and 0.22. During the former we are observing the post-AGB star at a time when the line of sight passes over (and under) the secondary star. The V-band continuum brightness of the system reaches its maximum around this phase thanks to more favorable scattering conditions for continuum radiation arising from the primary, as shown by Waelkens et al. (1996). In Figure~\ref{ha17-13} we compare the observed H$\alpha$ emission profiles derived from our spectra \#17 (phase 0.580) and \#13 (phase 0.216), which correspond closely to superior and inferior conjunction, respectively. These two spectra were obtained within a time interval of less than four months during the same orbit. We observe a strong, broad, blue-shifted ($\sim -80 \mathrm{km\ s}^{-1}$) absorption feature depressing the continuum at superior conjunction (\#17), while at inferior conjunction (\#13), when we are viewing the system over and under the poles of the post-AGB star, we observe the broad H$\alpha$ emission plateau to be quite symmetrical by comparison, without the presence of any absorption. The interpretation is straight forward: the fast bipolar outflow, which produces a spectroscopic signature in the form of a relatively symmetric broad emission plateau during most orbital phases, produces a strong, blue-shifted absorption of the post-AGB star's continuum at the time of  superior conjunction, when the post-AGB star is observed over the poles of the secondary. This indicates that the origin of the fast bipolar outflow is located in the immediate vicinity of the secondary star in the system. We suggest that a bipolar jet is being launched from an accretion disk surrounding the secondary in a direction perpendicular to the orbital plane of the system. The maximum absorption in the jet is caused by matter flowing at a speed of approximately 100 km s$^{-1}$, after having been corrected for the inclination angle of 35\degr. An extended, blue absorption wing can be seen to outflow velocities in excess of  500 km s$^{-1}$, corresponding to the emission wings seen in Figure 4. We suggest that this jet provides the primary shaping force of the present bipolar nebula. The observation of the H$\alpha$ line profile at the phase of inferior conjunction (\#13) supports  this conclusion. At this phase we are observing the secondary in the background, and we are seeing both red-shifted and blue-shifted emission components to either side of the comparatively stationary central emission spike, with comparable strengths. There appears to be nothing above or below the primary which absorbs any of this emission.

\subsubsection{Jet Characteristics as a Function of Phase}
The phase of inferior conjunction affords the most unimpeded view of the jet associated with the secondary, without any noticeable reduction in the H$\alpha$ line flux due to absorption. We used our H$\alpha$ emission profiles, after correction for the stellar absorption line profiles described earlier and after subtraction of the narrow core emission spike, to evaluate the strength of absorption of stellar light by the jet as a function of phase by comparing integrated H$\alpha$ line fluxes with that observed at the phase of inferior conjunction. We integrated the total emission flux in each of the normalized spectra after continuum subtraction, then re-normalized for the phase-dependent variation in the continuum intensity (Waelkens et al. 1996), and subtracted each result from the integrated emission flux found in spectrum \#13, which corresponds to the phase of inferior conjunction. The result is shown in Figure~\ref{jetabs}, where the two vertical lines indicate the phases of inferior (0.22) and superior (0.59) conjunction, respectively. The vertical scale measures the strength of H$\alpha$ absorption relative to that at the phase of inferior conjunction. The peak in absorption does indeed occur near the phase of superior conjunction, when the viewing line-of-sight intersects the jet. The absorption is highly blue-shifted, consistent with a view through rapidly out-flowing gas. Figure~\ref{jetabs} suggests that the phase dependence of this absorption is not symmetric around the phase of superior conjunction in a sense that might be expected, if part of the out-flowing gas is trailing behind the secondary as it moves along its orbit. The orbital speed of the secondary, while smaller than the outflow speed, is nevertheless not negligible, and a trailing effect could be produced, if the jet is launched over an extended region, such as the surface of an accretion disk, with spatially variable launch speeds.

We also used the stellar absorption-line corrected emission profiles for a study of the jet speed as a function of orbital phase. We measured the full width of the broad plateau emission at a standard intensity level (flux= 1.0, where the continuum-subtracted level is 0.0) and plotted the value expressed in km s$^{-1}$ units against phase in Figure 7. The plotted values are not corrected for the likely effective inclination of the system; if we adopt the inclination proposed by Waelkens et al. (1996), actual outflow velocities are higher by a factor of 1.22.
The vertical lines in Figure 7 indicate the phases of periastron (0.29) and apastron (0.79), respectively. The data show a very rapid rise in outflow speed to a maximum slightly before periastron, followed by a more gradual decline by about a factor two to a minimum around the time of apastron. The orbits of the binary components are known to be highly eccentric ($e = 0.37$; Men'shchikov et al. 2002). The data in Figure 7 indicate that the outflow speed is indeed phase-modulated, which could be the result of a phase-dependent mass-transfer rate from the primary to the secondary.  In view of the post-AGB nature of the primary and the relative proximity of the two components (Men'shchikov et al. 2002), mass transfer from the primary to the secondary occurs most likely through Roche lobe overflow. Prior to periastron the available volume of the Roche lobe of the post-AGB primary is greatly reduced, causing the mass transfer rate to be increased, while after periastron passage the Roche lobe again expands, reducing the rate of mass loss to the secondary. This variation in accretion power appears to be reflected in a phase-dependent variation in the outflow speed, as supported by the data shown in Figure 7. We address the issue of how a variable mass accretion rate may affect the outflow speed in \S5.5.

The transition from inferior ($\phi$ = 0.22) to superior (0.59) conjunction substantially overlaps with the transition from the phase of periastron (0.29) to apastron (0.79). While the effect of absorption by the out-flowing jets clearly dominates around the phase of superior conjunction, some of the phase dependence of the H$\alpha$ flux differences at other phases, shown in Figure 6, could also be a consequence of variations resulting from the phase-dependent accretion power variation suggested by the data shown in Figure 7.

\subsection{H$\alpha$ Light Curve}
Jura et al. (1997) attributed the narrow H$\alpha$ emission spike to a compact photo-ionized region coincident with the center of the RR. This conclusion was based, in part, on the narrow width of the feature, in part on the fact that the emitting volume is at rest within the RR and does not move with the orbital motion of either component of the central binary. They supported their conclusion with VLA continuum mapping of the RR at 3.6 cm, 2.0 cm, and 1.3 cm wavelength, which revealed a barely resolved central emission region with a spectrum characterized by a spectral index of 1.5, which the authors found to be consistent with free-free emission from ionized gas which becomes optically thin near a wavelength of 2 cm. They inferred the need for a source of Lyman-continuum photons with a luminosity of L$_{\mathrm{Lyman}} \sim [0.09 \times (d/330 pc)^2]\ \mathrm{L}\sun$ in order to maintain the observed radio flux.

One of the principal goals of the present investigation is to identify the source of this far-ultraviolet photon flux. Toward this aim, we fitted the central H$\alpha$ spike in our 17 absorption-corrected spectra with a narrow Gaussian that ignored the broad emission plateau, which had been attributed to gas associated with the jets described above. In Figure~\ref{hanarrowvki} we plot the resultant H$\alpha$ spike emission flux as a function of orbital phase, together with the V-band continuum brightness of HD~44179 from Waelkens et al. (1996) and the integrated KI 7699~\AA\ emission line flux from our spectra. The solid and dashed vertical lines indicate the phases of inferior (0.22) and superior (0.59) conjunction, respectively. The variations in the H$\alpha$ flux clearly support variability with a period equal to the orbital period, although small variations around a strictly periodic behavior appear to be present as well. The one data point clearly not consistent with the inferred period, located at phase 0.47, was measured in the first spectrum in this program obtained in 2001, while the bulk of our data were obtained seven and eight orbits later.  The most important clue provided by the H$\alpha$ data in Figure~\ref{hanarrowvki} is the fact that the H$\alpha$ flux peaks during the phase of inferior conjunction, when the secondary component and its surrounding environment are in a position of optimal visibility. This interpretation is supported by the fact that the V-band brightness of the HD~44179 system reaches its maximum at the phase of superior conjunction. The V-band flux is almost entirely due to the primary post-AGB component, and it reaches the position of optimal visibility at the phase of superior conjunction, as noted by Waelkens et al. (1996). This leads us to the conclusion that the source of the ionizing photons in the RR moves around the orbit with the secondary component. 

The H$\alpha$ and V-band light curves are normalized to the same flux scale. We note that the amplitude in the phase-dependent H$\alpha$ flux variation is about three times as large as the V-band amplitude. Since the orbit of the secondary component in HD~44179 is most likely smaller than the orbit of the photometric primary (see \S4.1), the larger H$\alpha$ emission amplitude suggests that the region of ionized gas responsible for this emission lies well outside the orbits of either star, but it responds directly to the position of the secondary within its orbit. At the phase of inferior conjunction, the far side of the circumbinary environment is directly illuminated by the secondary and its surroundings, while at the phase of superior conjunction the optimally visible region of H$\alpha$ emission on the far side of the orbit is blocked from receiving Lyman continuum photons from the secondary by the large body of the post-AGB primary. Given the diameter of the primary of $\sim 0.4$ AU and the orbital parameters of the system, the primary produces a shadow cone with an opening angle of about 25\degr\ devoid of ionizing radiation on the opposite circumbinary wall with respect to the secondary. This explains the large-amplitude variation in the flux of the H$\alpha$ core emission.

For comparison, we added the KI 7699~\AA\ emission flux data to Figure~\ref{hanarrowvki}. This emission is remarkably constant over all phases. This is not surprising, once one recalls that the first ionization potential of K is 4.32 eV. The primary post-AGB component  (T$_{\mathrm{eff}} \sim 7700$~K) emits copious amounts of photons of this energy and is therefore the dominant source for the photons responsible for the ionization and excitation of K. All parts within the central cavity of the RR are constantly being illuminated by the primary, independent of phase, and as a result the ionization of K is relatively constant everywhere. This is clearly not the case with hydrogen (I.P. = 13.6 eV), which cannot be ionized by the primary to any significant degree. We will explore the nature of the process producing the hydrogen-ionizing photon in the following sections.

\section{The Nature of the Secondary}
\subsection{Mass of the Secondary}
HD~44179 is a single-line spectroscopic binary with a mass function of 0.049 M$\sun$ (see Equ. (1)). The corresponding mass of the secondary is a function of the unknown mass of the primary and the inclination \emph{i} of the orbital plane. Men'shchikov et al. (2002) assumed a mass for the primary of 0.57 M$\sun$, near the lower end of the range of masses found for post-AGB stars (Bloecker 1995) and \emph{i} = 79\degr, which equals the estimated tilt of the axis of the RR relative to the line of sight. With these values, they arrived at an estimated mass of the secondary of 0.35 M$\sun$, and they argued that this object was a low-mass helium white dwarf of high temperature and sufficient far-ultraviolet luminosity to account for the observed compact HII region in the RR. Waelkens et al. (1996), however, argued convincingly that the ``effective'' inclination is much smaller than 79\degr, because their data indicated that the central star system in the RR is not being viewed directly but more likely via scattered light only, with an ``effective'' inclination of about 35\degr. In fact, the near-infrared high-resolution images of the central portions of the RR presented by Men'shchikov et al. (2002) support this claim that the actual stars in the RR are obscured by the circumbinary disk seen nearly edge-on, with stellar light visible only via scattering above and below the disk plane.

With an ``effective''  inclination of 35\degr\ and the identical value of the mass function for the system, the resulting mass for the secondary turns out to be larger than that of the primary. We will argue (\S5) that the mass-loss phase of the post-AGB star in the HD~44179 system is still ongoing and will assume with Waelkens et al. (1996) that its current mass may be as high  as 0.8 M$\sun$, near the upper limit of the mass range for post-AGB stars (Bloecker 1995). With this value and the ``effective'' inclination of 35\degr, the corresponding mass of the secondary star is 0.94 M$\sun$. These estimates for the masses of the post-AGB star and its secondary (combined mass 1.74 M$\sun$) are in excellent agreement with the combined central mass inferred by the observation of the orbiting velocities of the CO gas in the outer circumbinary disk by Bujarrabal et al. (2005), who derive a value for the central mass of 1.7 M$\sun$ for an assumed distance of 710 pc, the value for the distance found by Men'shchikov et al. (2002) and adopted by us. Thus, with a mass well above that of a typical white dwarf, it seems much more likely that the secondary is indeed an unevolved  low-mass main sequence star, as suggested by Waelkens et al. (1996). We can substantiate this conclusion by pointing to the observed value of the maximum speed of the bipolar outflow (\S3.1.1), which we have shown to be associated with the secondary. It is generally found that jet speeds launched in circumstellar accretion disks are of the order of the escape speed from the vicinity of the accreting object (Livio 1997). Our spectroscopic data suggest a maximum outflow speed of about 560~km s$^{-1}$ for the secondary in HD~44179, with an average speed for the bulk of the mass causing maximum absorption at the phase of superior conjunction of about 100 km s$^{-1}$. These speeds are  comparable to the observed outflow speeds from disks around T Tau stars (Eisloeffel \& Mundt 1998; Ferreira et al. 2006), while jets launched from the vicinity of white dwarf stars are faster by at least one order of magnitude or more (e.g. Rupen et al. 2008). The escape speed at the surface of a 0.94 M$\sun$ main-sequence star is about 600 km s$^{-1}$. We, therefore, conclude that the secondary star in HD~44179 is a main sequence star of sub-solar mass, implying an effective temperature of less than 6000 K. Thus, this star itself cannot be the source of the Lyman continuum responsible for producing the compact HII region in the RR, nor can it be the source of the far-UV radiation needed to excite the ERE (Witt et al. 2006) in the RR.

\subsection{Mass Accretion through Roche Lobe Overflow}
It appears most likely that the mass ejected from the vicinity of the secondary is obtained via accretion from the mass-losing post-AGB primary. Several arguments suggest that this accretion occurs through a Roche lobe overflow rather than through the much less efficient process of accreting matter from a slow, isotropic stellar wind leaving the primary. 

The first argument comes from the post-AGB star's size. Our model atmosphere calculation (\S3.1.1) produces the best fit of the H$\gamma$ line profiles for a star with a stellar radius of $3.1\times10^{12}$~cm. The average  radius of the ABG star's orbit (Men'shchikov et al. 2002) is $8.9\times10^{12}$~cm, slightly more than twice the star's estimated radius, if we employ the ``effective inclination'' of 35\degr of Waelkens et al. (1996). The two components in HD~44179 are roughly of equal mass (\S4.1), with the secondary being 18\% more massive, and thus possess Roche lobes of roughly equal size. Given the values of the average orbital radius and the estimated size of the AGB star, we conclude that the outer atmosphere of the post-AGB star extends all the way to the inner Lagrangian point of the system during most of its orbit, i.e. that it more than fills its Roche lobe at periastron.  Mass loss, therefore, will occur most readily by Roche lobe overflow. As the inter-component separation varies by about a factor two over the course of the orbit, the rate of Roche lobe overflow is expected to vary with orbital phase.

A second argument comes from the relatively high orbital eccentricity of the HD~44179 orbit ($e = 0.37$; Men'shchikov et al. 2002). Generally, tidal interactions in a close binary stars are expected to circularize the orbit; instead, binaries with AGB components stand out by exhibiting highly eccentric orbits. This led van Winckel et al. (1995) to suggest that strongly enhanced mass transfer through Roche lobe overflow at the time of periastron could lead to an increase in eccentricity, more than counterbalancing the tidal effect. Soker (2000) studied this process in detail and confirmed the suggestion of van Winckel et al. (1995). A more recent study by Bonacic Marinovic et al. (2008) arrived at very similar conclusions. 

Thirdly, mass accretion through Roche lobe overflow involves a much higher rate of transfer of specific angular momentum than is the case for accretion from an isotropic wind, making the formation of an accretion disk more likely (Soker 2008). Finally, our own data (Figure 7) show that the speed of the bipolar jets emanating from the secondary is strongly phase-dependent, reaching a maximum shortly before periastron and a minimum at apastron. The energy liberated by accretion onto the secondary can vary only as a result of a variable mass accretion rate.  As the two components move in elliptical orbits, the volume of the AGB star's Roche lobe decreases as the components approach periastron, leading to enhanced mass overflow. The process then reverses as the two components recede from each other after periastron passage.

\section{The Source of Lyman Continuum Radiation}
\subsection{The Central HII Region}
We adopt the results of Jura et al. (1997) concerning the required Lyman continuum luminosity of the HD~44179 system in order to explain the photo-ionization of hydrogen and helium (Kelly \& Latter 1995) and to power the observed radio continuum flux observed from the central region of the RR by Jura et al. (1997). At the adopted distance of HD~44179 of 710 pc (Men'shchikov et al. 2002) this Lyman continuum luminosity is 0.42 L$\sun$.  Although the AGB component in HD~44179 has a luminosity of about 6,150 L$\sun$, it is not hot enough to produce the amount of Lyman continuum flux required by the observed ionization of hydrogen. Furthermore, our study of the variation of the flux in the narrow H$\alpha$ spike as a function of phase clearly suggests that the source of ionization moves with the secondary component. Since the secondary itself is most likely an ordinary main-sequence star of even lower temperature than the primary (\S4.1), the Lyman continuum implied by existing data most likely comes from an accretion disk surrounding the secondary.

\subsection{Steady Accretion Disk Model}
Although the detailed physics of accretion disks is rather complex and will not be discussed in this paper, overall properties of an accretion disk such as the accretion luminosity, the radial temperature distribution and the resultant emission continuum can be readily estimated from basic physical principles for a steady disk model (Pringle 1981). In particular, this approach will allow us to test our previous conclusions, by permitting us to determine whether the mass accretion rate required for the production of the Lyman continuum luminosity of 0.42 L$\sun$ is in fact in agreement  with observed mass loss rates found in ABG stars. By assuming a main sequence star with sub-solar mass and radius  as the accreting body rather than a white dwarf, as proposed by Men'shchikov et al. (2002), we can also test this particular aspect of the system.

We adopt from Pringle (1981) the expressions for the disk luminosity produced as a result of steady accretion
\begin{equation}
\begin{split}
L_{disk} &= \int_{R_*}^{\infty} D(R) \cdot 2\pi R\ dR\\
	&=\frac{GM\dot{M}}{2R_*},
\end{split}
\end{equation}	
for the radial temperature distribution
\begin{equation}
T_s=\left\{ \frac{3GM\dot{M}}{8\pi R^3\sigma}\left[1-\left(\frac{R_*}{R}\right)^{1/2}\right]\right\}^{1/4},
\end{equation}
and for the continuous spectrum
\begin{equation}
S_\nu \propto \int_{R_*}^{R_{out}} B_\nu[T_s(R)] \cdot 2\pi R\ dR.
\end{equation}

In these expressions, $D(R)$ is the energy dissipation rate as a function of radius $R$ in the accretion disk surrounding the secondary due to viscosity, $M$ is the accreting star's mass, $R_*$ is its radius, and $\dot{M}$ is the mass accretion rate from the photometric primary onto the accretion disk surrounding the secondary. The maximum disk temperature in this model is obtained at a distance of $\frac{49}{36}\ R_*$ from the accreting star's surface. It is assumed that each radial mass element radiates as a local black body at the temperature indicated for that radius.

We note that the disk luminosity of the steady accretion disk model represents only half of the total available accretion energy (Pringle 1981). The fraction unaccounted for takes the form of kinetic energy in the interface region between the innermost disk and the stellar surface. Therefore, the portion of the accreting matter not ejected as part of the bipolar jet is expected to release energy similar in amount to the disk luminosity upon impact on the star. Radiation  from the expected impact shock region is most likely a major contributor to the Lyman continuum as well. Consequently, by focussing just on the far-UV continuum expected from the disk, we will estimate an upper limit to the required mass accretion rate.

In Figure~\ref{diskspectra}, we show sample disk spectra for mass accretion rates ranging from $1\times10^{-6}$ M$\sun$\ yr$^{-1}$ to $1\times10^{-4}$ M$\sun$\ yr$^{-1}$ for a main sequence star of 0.94 M$\sun$, calculated from Equ. (4). We notice that all spectra extend into the ultraviolet well below the Lyman limit at 91.2~nm. In each instance, this short-wavelength radiation is produced in the innermost, hottest region of the respective disk model. The range of mass accretion rates explored in this graph is representative of mass loss rates encountered among post-AGB stars (Bieging et al. 2006) and is not to suggest that the mass transfer rate in HD~44179 is variable to such a degree.

Figure~\ref{diskchars} summarizes the overall characteristics of the steady disk models as a function of the mass accretion rate over the same range of values as above. Being directly proportional to the mass accretion rate, the total disk luminosity varies approximately from $\sim 10$\ L$\sun$\ to $1000$\ L$\sun$. The maximum disk temperature varies from about 8000 K at the lowest mass accretion rate considered and rises to more than 24,000 K in the case of the highest mass accretion rate. The Lyman continuum luminosity constraint of Jura et al. (1997) is indicated in Figure~\ref{diskchars} by a dashed horizontal line. We calculated the Lyman continuum luminosities expected from our models by integrating the luminosity shortward of the Lyman limit for the model spectra shown in Figure~\ref{diskspectra}. We find that a steady disk model with a mass accretion rate of $\sim 2\times10^{-5}$\ M$\sun$\ yr$^{-1}$ matches the Lyman continuum constraint imposed by the Jura et al. (1997) observations. The maximum temperature in this disk is about 17,000~K.

The black-body assumption for the emerging spectra of the steady-disk model in the temperature range applicable to the HD~44179 accretion disk is probably not a very reliable approximation in the Lyman continuum range, where the opacity due to neutral hydrogen will be very high. Thus, it is likely that the emergent flux shortward of 91.2~nm wavelength will be depressed and redistributed to longer wavelengths (Kriz \& Hubeny 1986; Hubeny et al. 2000). Consequently, the maximum disk temperature and associated mass accretion rate needed to produce a specific Lyman continuum luminosity, as predicted by the steady-disk model, will be lower limits to the actual values. If we approximate the Lyman continuum spectrum of the accretion disk by that of a B-star atmosphere producing the same Lyman continuum flux as a black body at the lower temperature, an effective temperature near 23,000 K is required in the hottest region of the accretion disk. This increases the required mass accretion rate from about $2\times10^{-5}$ M$\sun$\ yr$^{-1}$ to $5\times10^{-5}$ M$\sun$\ yr$^{-1}$.

\subsection{Mass Accretion Rate}
The mass accretion rate of $2 - 5\times10^{-5}$\ M$\sun$\ yr$^{-1}$ implied by the Lyman continuum luminosity constraint for HD~44179 derived by Jura et al. (1997), assuming that the accreting secondary is a main sequence star of 0.94 M$\sun$, is in excellent agreement with observationally determined mass loss rates of post-AGB stars and stars found at centers of proto-planetary nebulae (Bieging et al. 2006; Ramstedt et al 2008). The agreement of these two rates implies that the transfer of mass from the post-AGB primary to the main-sequence secondary in HD~44179 must be highly efficient, i.e. through a Roche lobe overflow.

Our estimate of the current mass loss rate of the post-AGB component of HD~44179 is also in accord with the mass-loss history of HD~44179, reconstructed from the density distribution of the RR by Men'shchikov et al. (2002), who assumed a spherical outflow with constant velocity of 5~km~s$^{-1}$ as the origin of all observed parts of the RR. While this outflow velocity is certainly too low in comparison with that observed by us for the bipolar jet, they arrive at a maximum outflow age of about $5\times10^{4}$ yr for the RR, with implied recent mass loss rates ranging from values as high as $5\times10^{-2}$\ M$\sun$\ yr$^{-1}$ to values as low as $4\times10^{-9}$\ M$\sun$\ yr$^{-1}$ for different observed components of the RR. Our estimate of $2\times10^{-5}$ M$\sun$\ yr$^{-1}$ to $5\times10^{-5}$ M$\sun$\ yr$^{-1}$ is readily accommodated within this range.

\subsection{Is the Far-UV Output from HD~44179 Variable?}
Both the disk luminosity and temperature distribution in the steady accretion disk model (Equs. 2 and 3) are directly proportional to the mass accretion rate. In the case of a mass-exchanging binary with a highly eccentric orbit such as HD~44179, we expect the mass transfer rate to be variable (Regos et al. 2005; Sepinsky et al. 2007a, 2007b). How likely is it that a periodically variable mass accretion rate with P = 318 d as in HD~44179 produces a correspondingly variable far-UV radiative output? 

The limited observational data on the far-UV flux from HD~44179 (Sitko et al. 2008) covers only the spectral range accessible to spectrometers on IUE and HST; within observational uncertainties these observations suggest a constant far-UV flux level. There are at least two strong reasons for this observed fact. 

One, the accretion-powered far-UV and Lyman continuum radiation is expected to emerge from the hot, innermost region of the disk and from potential impact shock regions on the stellar surface. The theory for time-dependent accretion disks developed by Lightman (1974) allows us to estimate the time scale for radial drift of material accreted at the outer periphery of the disk. For representative parameters for a disk around a 0.94 M$\sun$ main-sequence star, this time scale is of the order of 500 years. It is, therefore, almost certain that short-term modulations with P = 318 d  of the mass accretion rate at the outer disk periphery will be fully smoothed out by the time the material arrives in the innermost disk region. Therefore, the far-UV emission from the disk as well as potential Lyman continuum radiation from the stellar surface are not expected to exhibit short-term periodic fluctuations.

A second, contributing reason for the observed constancy of the UV flux from HD~44179 comes from the fact that the far-UV radiation from the photosphere of the post-AGB component is still providing the dominant contribution to the flux in the spectral range accessible to IUE and HST. Our atmosphere model discussed in \S3.1.1 predicts a flux near 2000 \AA\ that is about one order of magnitude larger than the continuum flux expected from the steady accretion disk model for a mass accretion rate of $2\times10^{-5}$ M$\sun$\ yr$^{-1}$. 

The range of variation in the mass accretion rate between periastron and apastron is difficult to estimate. It depends on the exact orbital phase at which the volume of the post-AGB primary exactly fills its Roche lobe as well as the density scale height in the envelope of the primary. Furthermore, the expected decline in the direct mass transfer rate from primary to the disk of the secondary upon increasing separation between the two components in HD~44179, while moving from periastron to apastron, will likely be compensated, at least in part, by increasing mass accretion from the circumbinary environment as the Roche lobes expand during the transition from periastron to apastron. 

\subsection{What Drives the Bipolar Outflow?}
Our measurements of the width of the plateau emission profile attributed to the bipolar jet shown in Figure 7 suggest a strong modulation of the jet characteristics with orbital phase. The data can be interpreted in terms of a variation in the jet speed or a variation in the mass outflow rate, or both. Either way, in light of the expected steadiness of the radiative output from the disk, this result suggests that the outflow is not radiatively driven. We can confirm this by comparing the momentum carried by the outflow with the radiative momentum available from a 300 L$\sun$ accretion disk expected from a mass accretion rate of $2\times10^{-5}$ M$\sun$\ yr$^{-1}$. Assuming that 10\% of the accreted mass (Papaloizou \& Pringle 1977) leaves the system as a bipolar jet with a characteristic speed of 100 km s$^{-1}$, the jet momentum exceeds the radiative momentum available for radiation pressure by a factor of about 50. The observed jets cannot be driven radiatively.  It appears likely, therefore, that the launching of the jets and their collimation are a result of  magneto-hydrodynamical processes (e.g. Pudritz \& Norman 1983; Casse \& Keppens 2004).

The variation of the jet characteristics with orbital phase is probably related to the width of the jet. The strength of maximum jet absorption at the phase of superior conjunction (Figure 4a), when compared with the available flux in the predicted wings of the stellar H$\alpha$ profile (Figure 3), suggests that the jet is optically thick at the blue-shifted wavelength of maximum absorption and that it covers the entire stellar disk of the post-AGB star in projection. This implies that the jet is at least 0.4 AU wide, comparable to the size of the accretion disk around the companion star, and therefore is directly affected by the variations in the mass accretion rate.

\section{Discussion and Conclusions}
\subsection{Binaries in Proto-Planetary Nebulae}
Ongoing research into the shapes and shaping mechanisms of bipolar planetary nebulae and proto-planetary nebulae (Balick \& Frank 2002; Soker and Livio 1994; Nordhaus \& Blackman 2006; Akashi 2007) suggests that there may be a number of distinctly different mechanisms at work in different objects. It is also possible that different mechanisms may dominate the morphological evolution of a single object during different phases of its evolution (e.g. Waters et al. 1998). Our current result concerning the presence of a bipolar jet associated with the secondary star in HD~44179, which appears to be energized by mass accretion into a circumstellar disk through Roche lobe overflow on part of the post-AGB primary is illustrated as an artistic representation in Figure 11. We believe HD~44179 to be a system where a binary has survived an earlier common-envelope phase, which produced the outer circumbinary disk, within which the remaining semi-detached binary is now engaged in mass transfer. This particular configuration may be the result of a specific set of initial conditions and it may also be short-lived. This picture should, therefore, not be generalized too far and should not be applied indiscriminately to all bipolar nebulae (Allen et al. 1980).  Our data support the model proposed  by Morris (1987) for the RR and developed later by Soker (2000, 2005). This, however, may only explain the cause of the \emph{present} outflow from the central cavity of the RR. We emphasize that the accretion disk around the secondary discussed here is much smaller than and physically distinct from the much larger circumbinary disk, which has been resolved in high-resolution near-infrared (Tuthill et al. 2002) and optical images (Osterbart et al. 1997; Cohen et al. 2004; Vijh et al. 2006). The presence of this outer disk, which is in Keplerian revolution around the central binary (Bujarrabal et al. 2005), very likely assists in restraining any current mass outflow into bipolar directions. This outer disk may have had its origin during an earlier common-envelope phase, one of the mechanisms discussed by Nordhaus \& Blackman (2006). Both the current mass ejection process in the RR as well as a possible earlier common-envelope phase depend critically upon the binarity of HD~44179, and this particular aspect of the RR appears to be rather common among bipolar nebulae (e.g. Yamamura et al. 2000; de Ruyter et al. 2006; van Winckel et al. 2006). A recent \emph{GALEX} search for ultraviolet excesses among AGB stars by Sahai et al. (2008) revealed detectable excesses in about 43\% of their sample, suggesting the presence of either hot companions or accretion upon low-mass secondaries. This is about three times the confirmed binary frequency among central stars in planetary nebulae (De Marco 2006), and goes quite far in supporting the contention that binarity is required for the production of planetary nebulae.

\subsection{Excitation of Extended Red Emission }
The RR appears to be unique among known bipolar proto-planetary nebulae for the remarkably high intensity of extended red emission (ERE) (Schmidt et al. 1980; Witt \& Boroson 1990), which is responsible for the X-shape morphology of the nebula at red wavelengths. Detailed structural studies of the RR (Schmidt \& Witt 1991;  Cohen et al. 2004;  Vijh et al. 2006) have found the ERE in the RR to be most intense on nebular surfaces which are illuminated directly by the central source, consistent with an ERE excitation mechanism involving photo-excitation by far-ultraviolet photons with energies E $>$ 10.5 eV (Witt et al. 2006). A potentially embarrassing problem for this scenario is the fact that neither the post-AGB primary nor the main-sequence secondary in HD~44179 emits nearly enough radiation with E $>$ 10.5 eV to account for the total number of ERE photons observed from the RR. The suggestion that the secondary might be a hot He white dwarf advanced by Men'shchikov et al. (2002) appeared to be a welcome solution to this problem, but our current observations do not support this proposal. Now, the hot, innermost region of the proposed accretion disk with a spectrum as shown in Figure~\ref{diskspectra} replaces the role of the white dwarf companion in providing the needed far-ultraviolet photons for the excitation of the ERE. Witt et al. (2006) estimate the total ERE luminosity in the RR, corrected for extinction, as L$_{\mathrm{ERE}} \sim 10$ L$\sun$, and they proposed a two-stage process of creation of the ERE carrier through ionization of a precursor, followed by optical pumping with near-UV/optical photons to explain the production of ERE. The luminosity of far-UV photons (E $>$ 10.5 eV) obtained from the disk for a mass accretion rate of $2\times10^{-5}$ M$\sun$\ yr$^{-1}$ shown in Figure~\ref{diskchars}, with a spectrum as shown in Figure~\ref{diskspectra}, would be higher than the luminosity of Lyman continuum photons of E $>$ 13.6 eV  and sufficient to accomplish the critical first stage of this process, leaving the main task of optical pumping of the ionized ERE carrier to the post-AGB primary with its comparatively large near-UV/optical luminosity. In order to reproduce the conditions leading to the high ERE luminosity observed in the RR, apparently three conditions must be met: a powerful source of far-ultraviolet photons for the production of the ERE carriers, requiring a high mass-accretion rate onto an accretion disk surrounding the secondary; a still more powerful source of near-UV/optical photons for the optical pumping of the ERE carriers, in our case the HD~44179 post-AGB primary; and a carbon-rich chemistry with C/O $>$ 1. There is new, strong evidence (Lagadec \& Zijlstra 2008; Woitke 2006) showing that among AGB stars of luminosity L $< 10^4$ L$\sun$ and low metallicity only C-rich stars achieve mass-loss rates of $10^{-5}$ M$\sun$\ yr$^{-1}$ or above. These are all conditions that apply to HD~44179 and may point to the reason for the apparently unique prominence of ERE in the nebula associated with this star.

\subsection{Conclusions}
We summarize the findings of this investigation as follows:
\begin{enumerate}
\item RV measurements spanning 8 orbits during the years between 2001 and 2008 confirmed the orbital period of 318 days originally published by Waelkens et al. (1996) and constrain the uncertainty to $\pm$1 day. They do not support the period of 322 days reported later by Men'shchikov et al. (2002).
\item The emitting gas at the center of the RR characterized by narrow emission lines, e.g. by the narrow H$\alpha$ core and the KI 7699~\AA\ line, does not participate in the orbital motions of the binary components in HD~44179. Instead, the narrow-line gas is nearly at rest with respect to the system. This agrees with earlier results by Hobbs et al. (2004).
\item The complex H$\alpha$ emission line profile undergoes systematic changes that are periodic with the orbital period of 318 days. Five separate components are found to contribute to the H$\alpha$ profile: (a) a narrow (18 km s$^{-1}$ FWHM) emission spike at rest with respect to the RR system; (b) a broad emission plateau (FW $\sim 200$ km s$^{-1}$) moving with the orbital motion of the secondary while simultaneously changing its width from a maximum near periastron to a minimum near apastron; (c) weak, extended emission wings indicating maximum outflow velocities of $\sim560$ km s$^{-1}$; (d) a modulation of (a) and (b) by an underlying H$\alpha$ absorption line arising in the atmosphere of the post-AGB primary, with the orbital period of 318 days; and (e) a strongly blue-shifted H$\alpha$ absorption appearing near the phase of superior conjunction, when the lines of sight from the observer to the post-AGB primary pass over the secondary.
\item We interpreted these spectroscopic observations as indicating that the fast bipolar outflow in the RR originates in the vicinity of the secondary star. The speed of the outflow as well as a re-assessment of the likely mass of the secondary in light of the viewing geometry of HD~44179 proposed by Waelkens et al. (1996) suggest that the secondary most likely is a main-sequence star of 0.94 M$\sun$. The energy and momentum seen in the fast outflow appear to be generated by mass accretion from the mass-losing post-AGB primary onto the secondary. Our interpretation of the outflow supports the early model of Morris (1987). Our conclusions concerning the nature of the secondary support an earlier suggestion by Waelkens et al. (1996) but conflict with Men'shchikov et al. (2002), who proposed a secondary in the form of a low-mass helium white dwarf.
\item We found the H$\alpha$ spike emission flux to vary with the orbital period. The highest flux was observed near the phase of inferior conjunction for the primary, when the secondary is on the far side of its orbit. This suggests that the source of the ionizing radiation in HD~44179 travels with the secondary. The large amplitude of the H$\alpha$ spike emission flux as a function of phase can be understood in view of the large shadowing effect for ionizing radiation caused by the large disk of the primary.
\item We estimated the mass accretion rate needed to produce the disk luminosity and temperatures to generate the Lyman-continuum luminosity required to maintain the central compact HII region in the RR, detected at radio-continuum frequencies by Jura et al. (1997). We employed a steady accretion disk model and found that a mass accretion rate of $\sim2\times10^{-5}$ M$\sun$\ yr$^{-1}$ is sufficient to produce the needed photon flux, although more realistic modeling of the Lyman continuum portion of the disk spectrum may push the mass accretion rate higher, to $\sim5\times10^{-5}$ M$\sun$\ yr$^{-1}$. Such a disk is expected to have a luminosity of about 300 L$\sun$, spread out over a broad spectrum ranging from the Lyman continuum through the optical and the near infrared, as illustrated in Figure 9. The predicted disk luminosity is about 5\% of the luminosity of the post-AGB primary.
\item The estimated mass accretion rate for the secondary is in excellent agreement with observationally determined mass-loss rates of post-AGB stars in the literature. This suggests that the mass transfer in the HD~44179 system occurs with very high efficiency through Roche lobe overflow. This mechanism is supported also by the estimated size of the primary, suggesting that it fills its Roche lobe; the high orbital eccentricity, which can be maintained by enhanced mass overflow at the phase of periastron; and by the observation that shows that the speed of the bipolar outflow originating near the secondary decreases by close to a factor two between the time of periastron and apastron. This suggests that the liberated accretion energy varies with the orbital period in response to a mass overflow rate, which is highest near periastron. 
\item We show that the far-UV luminosity of the accretion disk is expected to be unaffected by the modulation in the mass accretion rate with orbital phase, in view of a long radial drift time scale. The momentum carried by the jets in HD~44179 greatly exceeds the radiative momentum available from the disk's luminosity. We conclude that the jets must be driven and collimated by magneto-hydrodynamical processes occurring over a large extent of the accretion disk.
\item A unique feature of the RR is the intense extended red emission (ERE), which requires far-UV radiation for its initiation, in addition to the presence of large carbonaceous molecules (Witt et al. 2006). The proposed accretion disk around the secondary generates enough far-UV radiation to power the ERE observed in the RR. Other systems at a similar evolutionary stage may have a lower mass transfer rate, generating insufficient UV radiation (Figure~\ref{diskchars}) or lack the required carbon-rich chemistry and thus lack the preconditions for ERE, making the RR relatively unique in this regard.
\end{enumerate}

\acknowledgments
ANW acknowledges a stimulating discussion about HD~44179 with Michael Sitko and Larry Bernstein. We received constructive suggestions concerning the characteristics of the accretion disk and the jet from Arieh Koenigl and Vikram Dwarkadas, for which we are grateful. Also, we gratefully acknowledge Steven Lane, who contributed his talent in producing the image of HD~44179 shown in Figure 11. Finally, we express our gratitude to the referee, Michael Sitko, for constructive criticism and positive suggestions, which led to significant improvements in the final version of this paper. ANW and UPV acknowledge support from the US National Science Foundation through grant AST0606756 to the University of Toledo.


\clearpage

\clearpage
\begin{figure}
\centering{
\includegraphics[angle=90,width=5.5in]{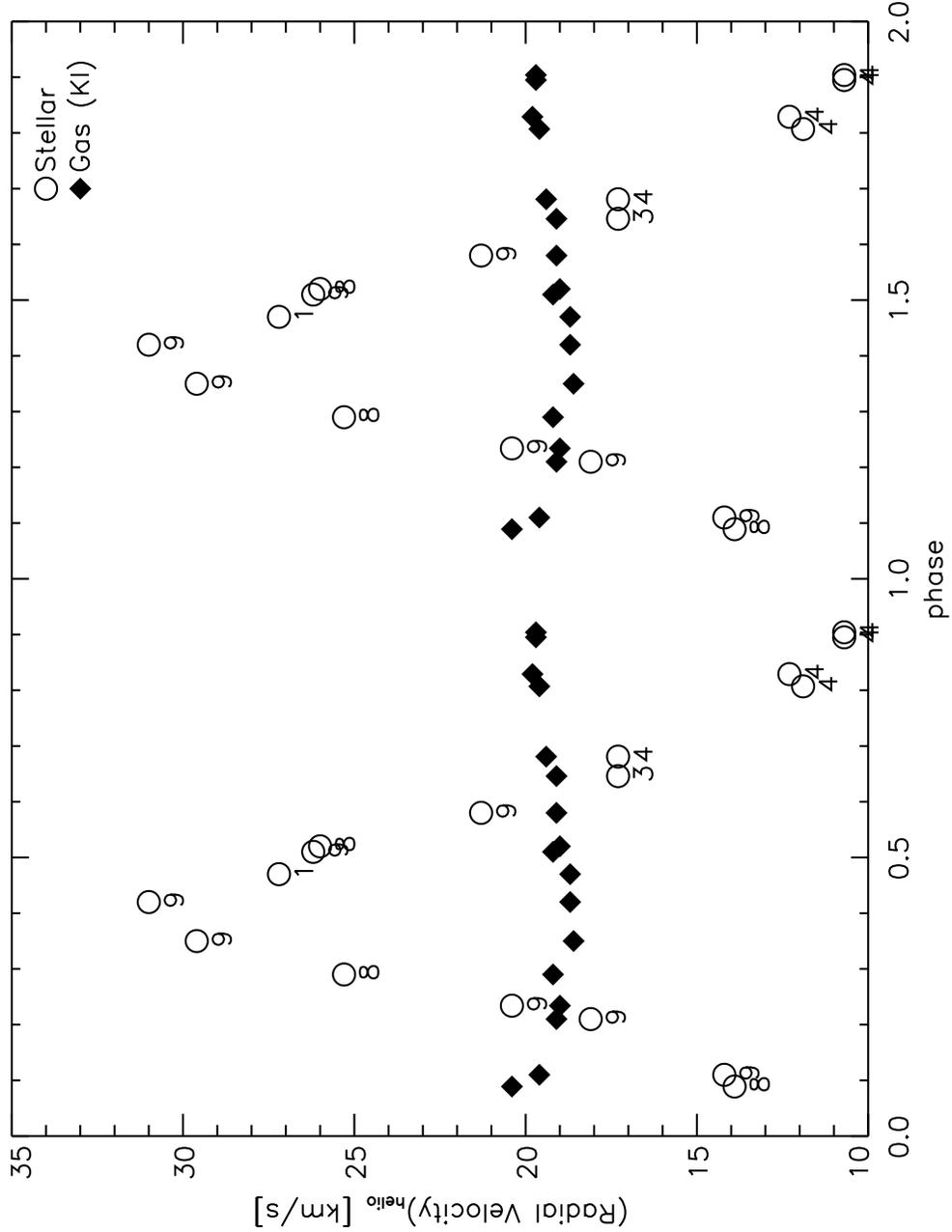}
\caption{The observed heliocentric stellar radial velocity of HD~44179 (open circles) are shown as a function of orbital phase. Data from eight orbits have been reduced with an adopted orbital period of 318 days. Digits associated with individual measurement points refer to the corresponding orbit (Table~\ref{t-obs}).  Also shown is the heliocentric radial velocity of the narrow KI 7699~\AA\ (filled diamonds) emission arising in the central HII region filling the inner space of the circumbinary disk associated with the HD~44179 system.All radial velocity measurements shown here are subject to random errors of $\pm$0.4~km~s$^{-1}$.\label{radialv}}
}
\end{figure}

\clearpage
\begin{figure}
\centering{
\includegraphics[width=7in]{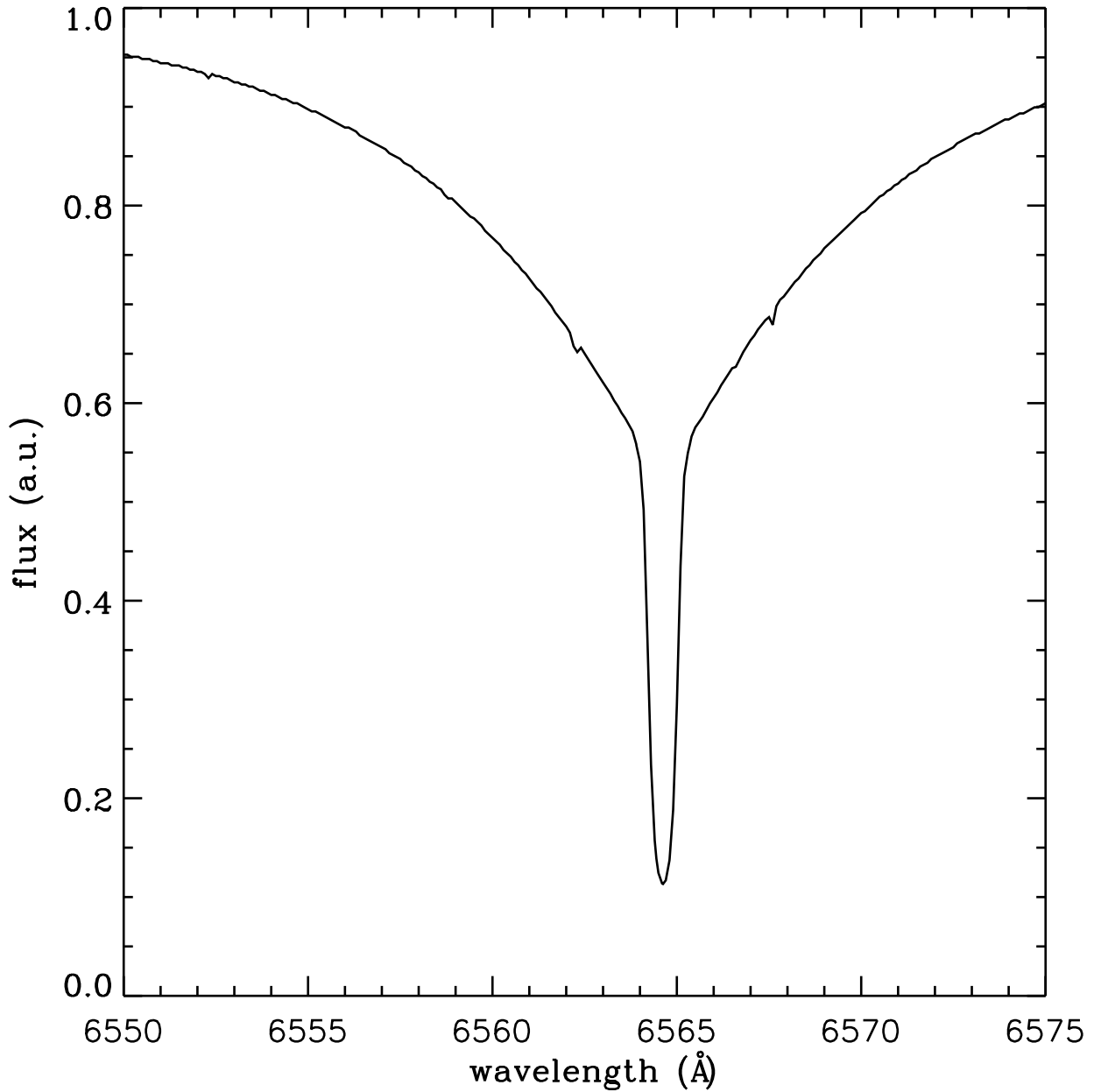}
\caption{The adopted stellar H$\alpha$ absorption line profile shown here was derived from the Phoenix NLTE stellar atmosphere code, using the observed stellar H$\gamma$ profile as a constraint. The adopted parameters are T$_{\mathrm{eff}} = 7700$~K, log g = 1.10, [Z/H] = -3.5, [CNO/H] = -0.5 and a stellar mass of 0.8 M$\sun$. The stellar radius is $3.1\times10^{12}$~cm, resulting in a bolometric luminosity near 6,150 L$\sun$. The flux scale is normalized to unity in the distant continuum and is expressed in arbitrary units (a.u.).\label{hamodel}}
}
\end{figure}

\clearpage
\begin{figure}
\centering{
\includegraphics[width=7in]{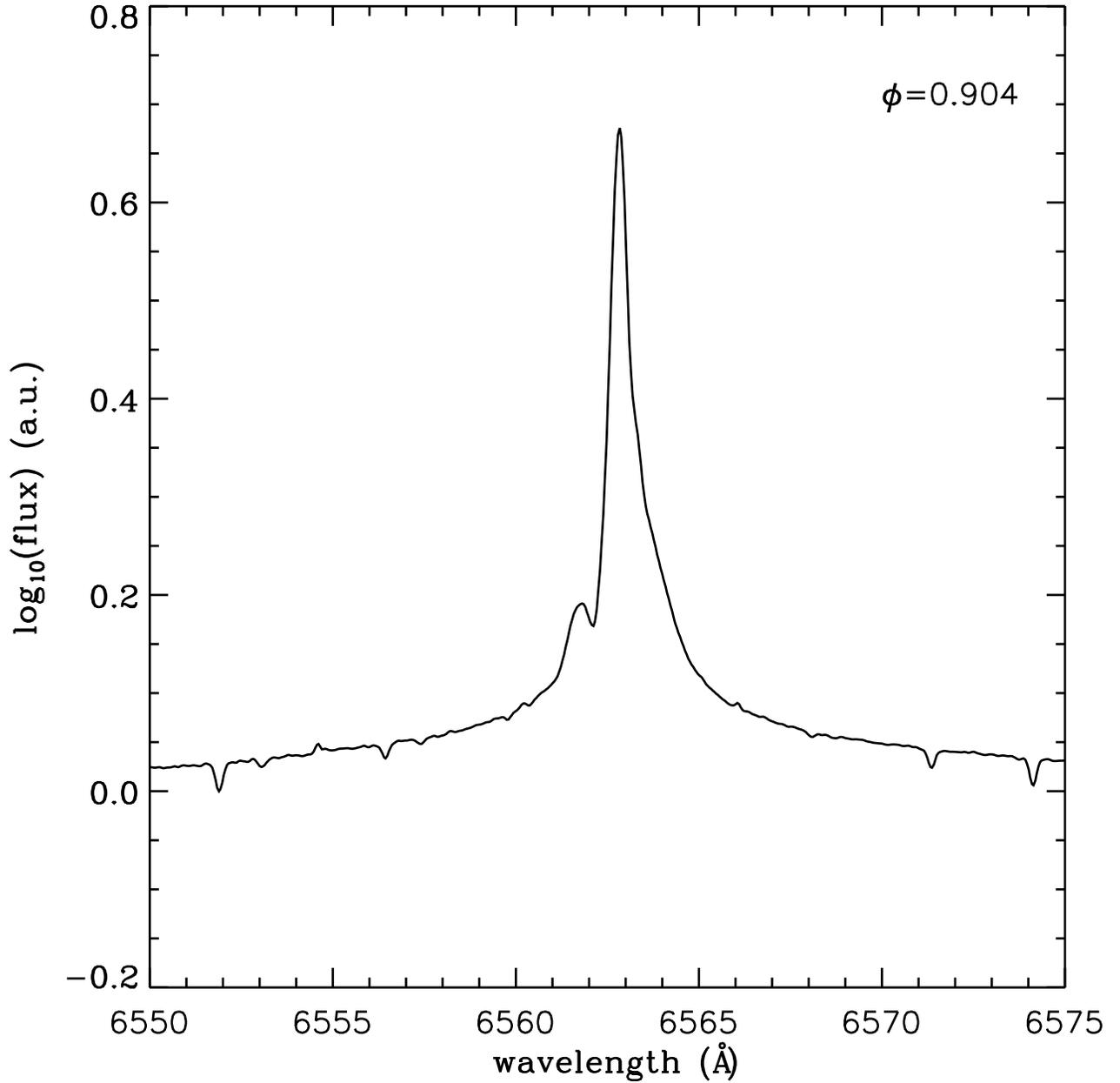}
\caption{H$\alpha$ emission line profile of HD~44179 at phase $\phi$= 0.904, corrected for stellar contribution in the form of the underlying stellar absorption line profile shown in Figure~\ref{hamodel}. Before subtraction, the predicted stellar H$\alpha$ absorption line profile was blue-shifted by 10.0 ~km~s$^{-1}$ relative to the emission line profile. The residual absorption dip can be accounted for through an insufficient width of the predicted model profile. The logarithmic flux scale is normalized to zero far from the line center and is expressed in arbitrary units (a.u.).\label{em17}}
}
\end{figure}

\clearpage
\begin{figure}
\centering{
\includegraphics[width=3in]{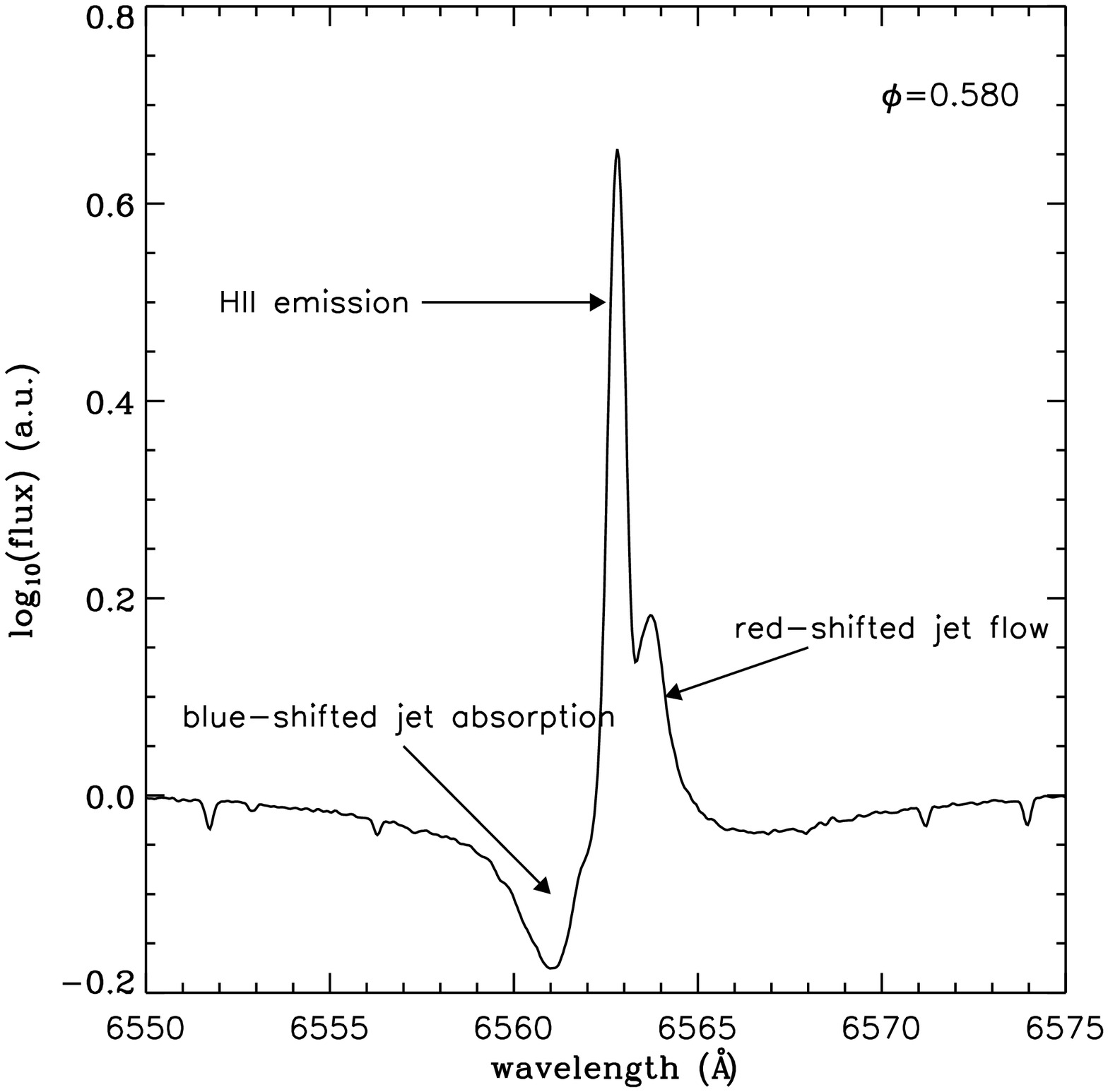}\includegraphics[width=3in]{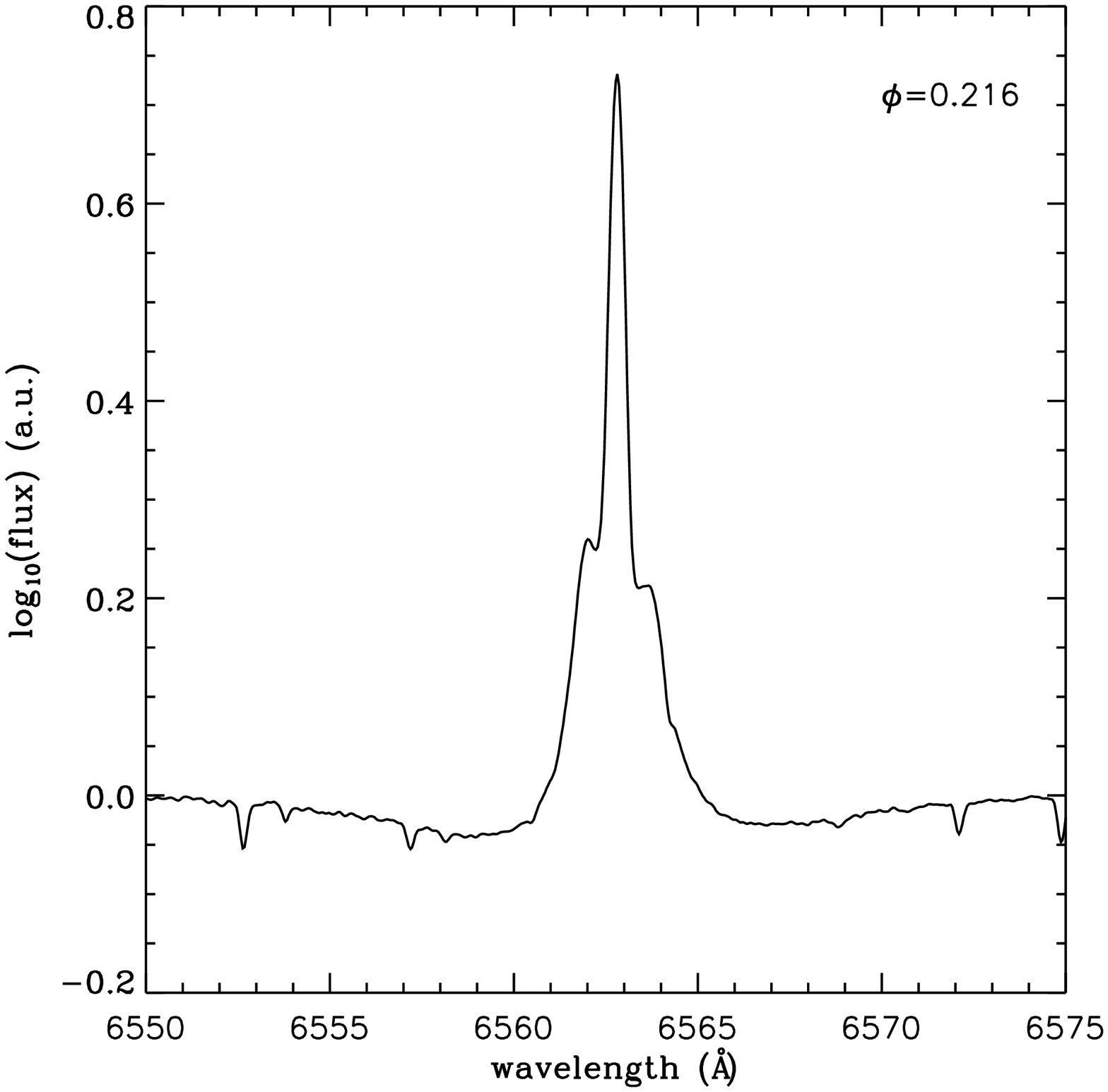}
\caption{(a) Observed H$\alpha$ emission line profile of HD~44179 at phase $\phi$= 0.58, closest to the orbital phase of superior conjunction, when the luminous primary star is viewed along lines of sight passing over the poles of the secondary companion. Note the appearance of a strong, blue-shifted absorption feature. (b) Observed H$\alpha$ emission line profile of HD~44179 at phase $\phi$= 0.216, closest to the orbital phase of inferior conjunction, when the primary component is between the secondary and the observer. Note that the H$\alpha$ emission consists of a narrow emission core and a broad , nearly symmetrical plateau of blue- and red-shifted emission. Normalization and flux scale are identical to those in Figure~\ref{ha7-15}.\label{ha17-13}}
}
\end{figure}

\clearpage
\begin{figure}
\centering{
\includegraphics[angle=90,width=6in]{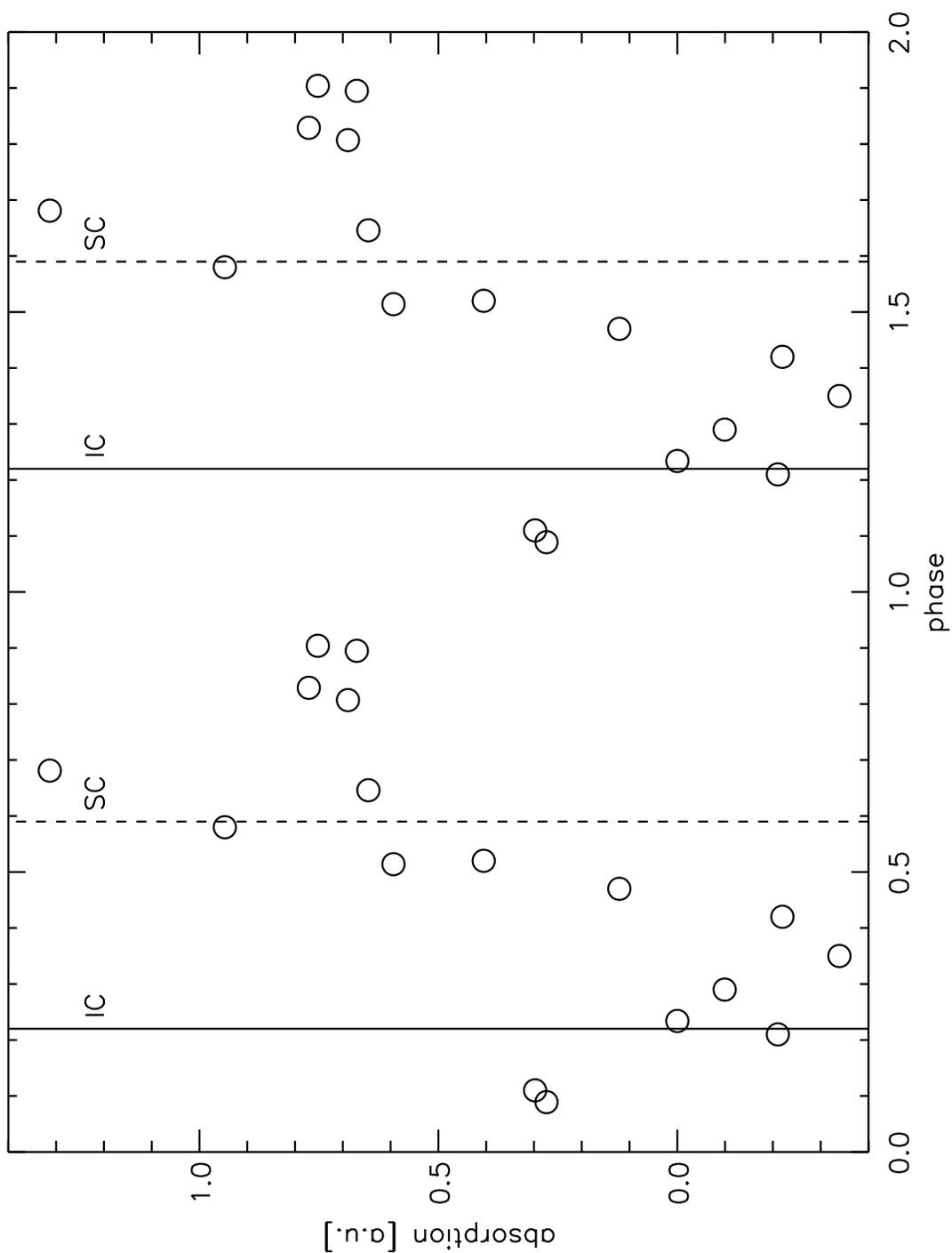}
\caption{Plot of the strength of the H$\alpha$ absorption due to the presence of the bipolar outflow in HD~44179 as a function of orbital phase. The plotted quantities are the differences in integrated H$\alpha$ emission flux between that measured at phase $\phi$= 0.216 (inferior conjunction) and that observed at other phases. The solid and dashed vertical lines indicate the phases of inferior (IC) and superior conjunction (SC), respectively. The absorption scale is in arbitrary units (a.u.) as explained in the text.\label{jetabs}}
}
\end{figure}

\clearpage
\begin{figure}
\centering{
\includegraphics[angle=90,width=6in]{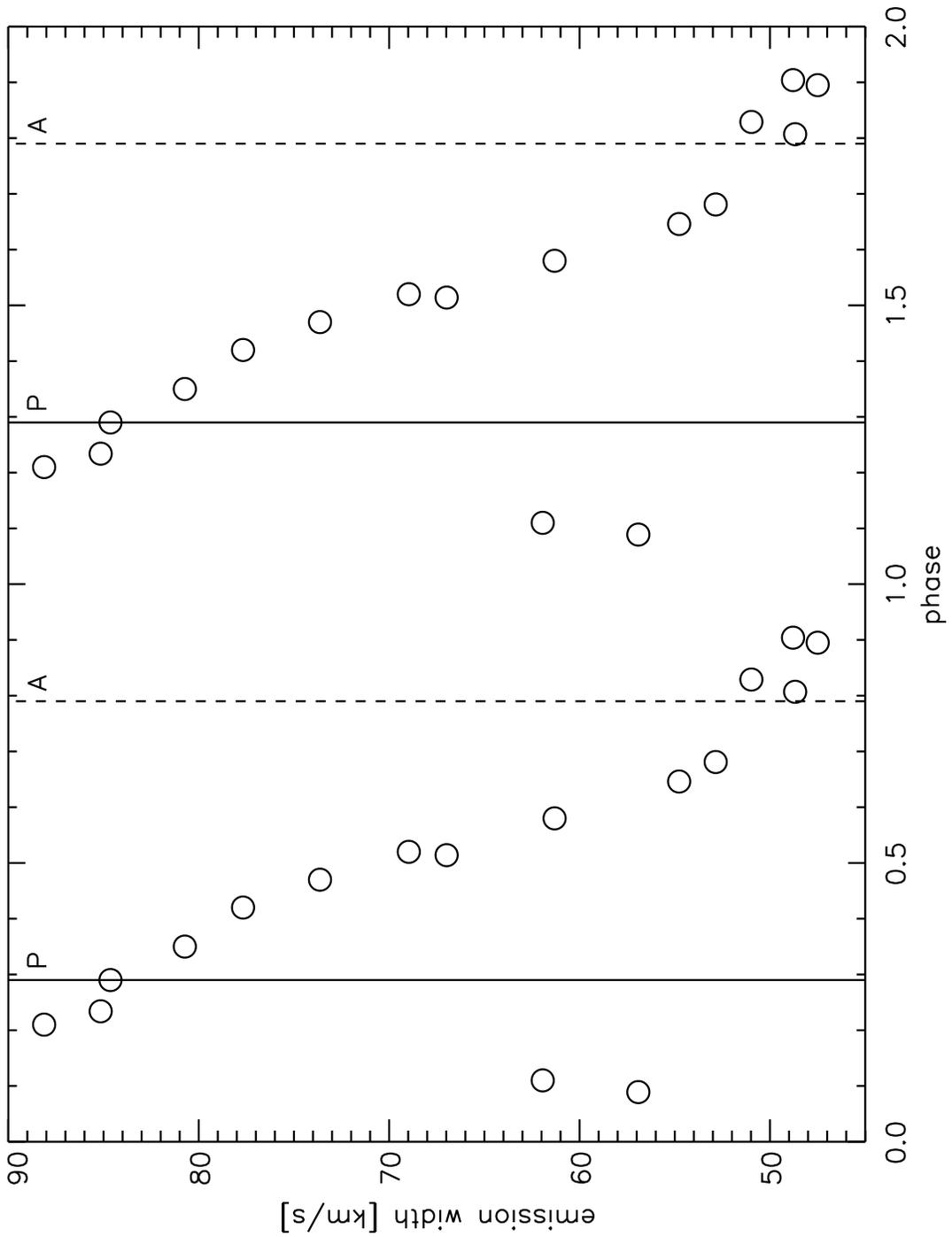}
\caption{Plot of the width of the H$\alpha$ emission plateau in HD~44179 as a function of phase at a normalized intensity level = 1, where zero intensity is the continuum-subtracted flux level far from the line center. The solid and dashed vertical lines represent the orbital phases of periastron (P) and apastron (A), respectively, in the eccentric ($e = 0.37$) orbit of HD~44179.\label{broadhaem}}
}
\end{figure}

\clearpage
\begin{figure}
\centering{
\includegraphics[angle=90,width=5.5in]{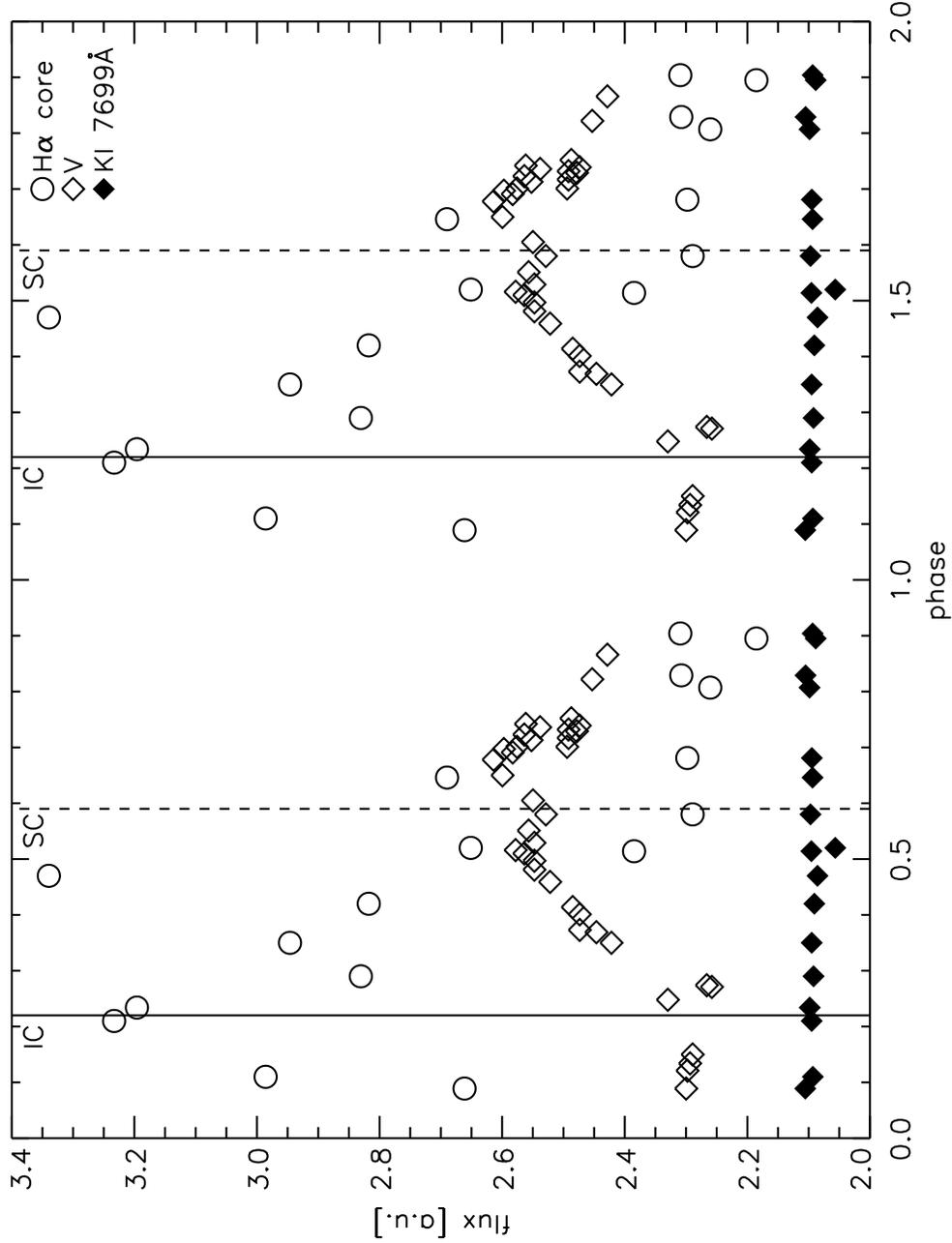}
\caption{Plot of the normalized light curves of HD~44179 as a function of orbital phase. The open circles represent the integrated H$\alpha$ emission line flux in the narrow spike, corrected for the stellar absorption (Figure ~\ref{hamodel}). The open diamonds trace the V-band continuum light curve of HD~44179 determined by Waelkens et al. (1996). The filled diamonds show the integrated emission in the KI 7699~\AA\  emission line as a function of phase. The solid and dashed vertical lines represent the phases of inferior and superior conjunction, respectively. Note that the H$\alpha$ emission peaks at the time of inferior conjunction (IC) while the V-band continuum displays a brightness peak at the time around superior conjunction (SC), when the post-AGB primary is in a position of optimum visibility.\label{hanarrowvki}}
}
\end{figure}

\clearpage
\begin{figure}
\centering{
\includegraphics[angle=90,width=6in]{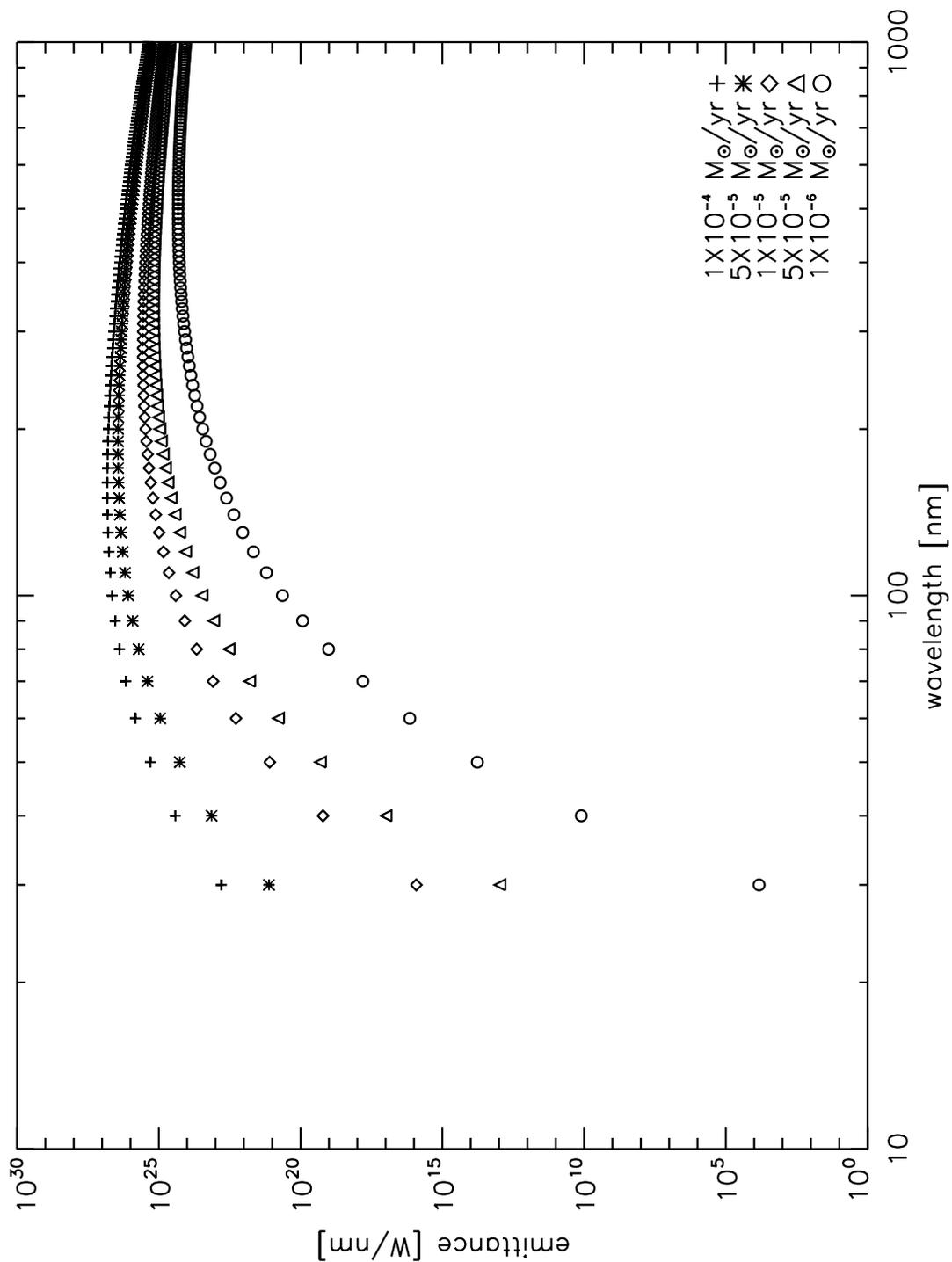}
\caption{Calculated disk emittance as function of wavelength of  accretion disks surrounding a $0.94$ M$\sun$ main sequence star for five mass accretion rates in the range from $1\times10^{-6}$  M$\sun$ yr$^{-1}$ to $1\times10^{-4}$ M$\sun$ yr$^{-1}$. This range is representative of mass loss rates encountered among post-AGB stars and is not to suggest that the mass loss rate of HD~44179 is variable to this degree.\label{diskspectra}}
}
\end{figure}

\clearpage
\begin{figure}
\centering{
\includegraphics[angle=90, width=5.5in]{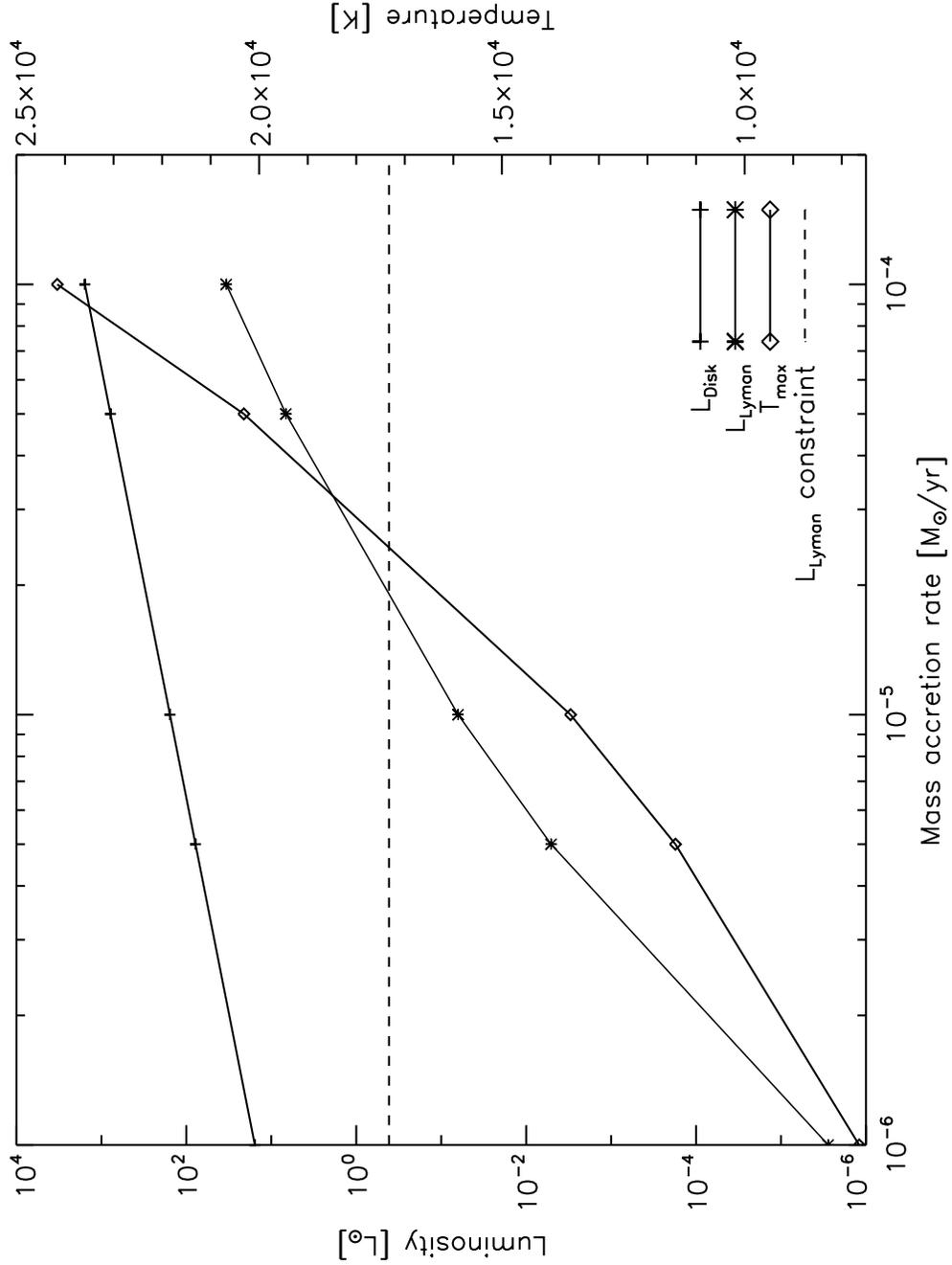}
\caption{Calculated total disk luminosity, maximum disk temperature, and Lyman continuum luminosity as a function of the mass accretion rate for a steady disk model centered on a $0.94$ M$\sun$ secondary component. The dashed horizontal line represents the constraint for the Lyman continuum required to maintain the photionization of the inner HII region in the RR observed by Jura et al. (1997), corrected for an assumed distance of the RR of 710 pc. In the present model, a mass accretion rate near $2\times10^{-5}$ M$\sun$ yr$^{-1}$ produces the required Lyman continuum from the inner disk region with a maximum disk temperature of about 17,000 K.\label{diskchars}}
}
\end{figure}

\clearpage
\begin{figure}
\includegraphics[width=6.5in]{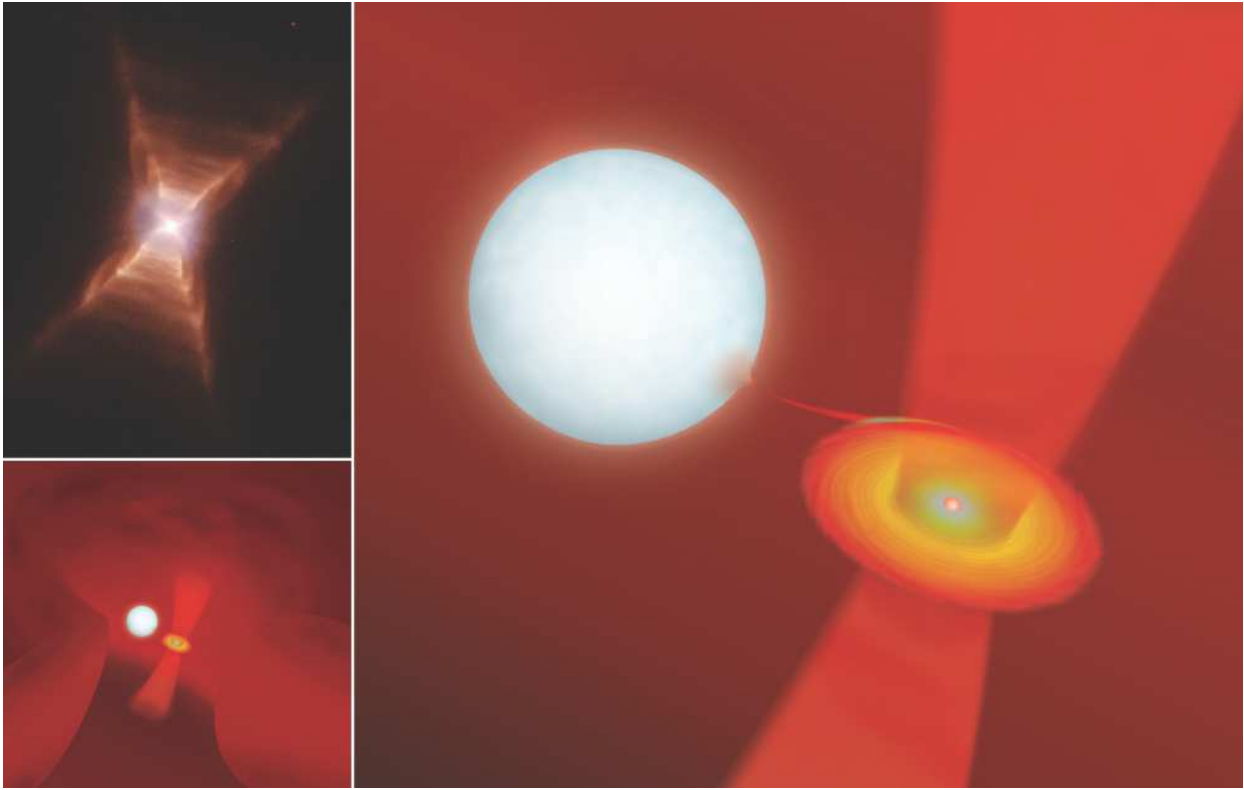}
\caption{This figure represents aspects of the RR on three different scales. The upper left insert is an HST color composite from the work of Cohen et al. (2004). The vertical extent along the bipolar axis of the RR shown in this image is about 15,000 AU at a distance of 710 pc. The main figure is an artist's rendition by Steven Lane of the possible appearance of the HD~44179 binary system, in accordance with the results and conclusions in the present paper. The linear scale of this image is about 1 AU. The lower left insert shows a cut-away depiction of how HD~44179 may be embedded in the central cavity of the circumbinary disk. The disk thickness is about 90 AU and the diameter of the central cavity is about 30 AU. The size of the binary within the cavity has been enlarged by about a factor ten to enhance its visibility at this scale. The slight curvature in the jet reflects the motion of its launch location with the secondary around the common center of mass of the HD~44179 system.\label{}}
\end{figure}

\clearpage
\begin{deluxetable}{rcrrrcrcrcc}
\tablecaption{\label{t-obs}}
\tablewidth{0pt}
\tablehead{\colhead{\#} & \colhead{Year} &\colhead{Month} & \colhead{Day} & \colhead{UT} & \colhead{HJD}  & \colhead{$\Delta$HJD}  & \colhead{Phase\tablenotemark{a}}  & \colhead{$\Delta$V$_{helio}$}  & \colhead{V$_{KI}$}	&\colhead{RV}}  
\startdata
1	&	2001	&	2	&	6	&	4:21:46	&	2451946.68531	&	0.00 	&	1.470	&	-16.44	&	18.7	&27.2\\
2	&	2002	&	12	&	31	&	7:15:39	&	2452639.80723	&	693.12	&	3.650	&	-1.63	&	19.1	&17.3\\
3	&	2003	&	11	&	24	&	8:25:03	&	2452967.85463		&	1021.17	&	4.681	&	13.93	&	19.4	&17.3\\
4	&	2004	&	1	&	3	&	8:15:46	&	2453007.84896	&	1061.16	&	4.807	&	-2.98	&	19.6	&11.9\\
5	&	2004	&	1	&	10	&	4:28:31	&	2453014.69072	&	1068.01	&	4.829	&	-5.60	&	19.8	&12.3\\
6	&	2004	&	1	&	31	&	4:58:31	&	2453035.71118	&	1089.03	&	4.895	&	-14.17	&	19.7	&10.7\\
7	&	2004	&	2	&	3	&	4:21:35	&	2453038.68538		&	1092.00&	4.904	&	-15.17	&	19.7	&10.7\\
8	&	2006	&	11	&	12	&	10:33:59	&	2454051.94356	&	2105.26&	8.090	&	17.66	&	20.4	&13.9\\
9	&	2006	&	12	&	28	&	5:36:38	&	2454097.73848	&	2151.05	&	8.234	&	-0.12	&	19.0	&20.4\\
10	&	2007	&	1	&	15	&	5:49:11	&	2454115.74694		&	2169.06&	8.291	&	-8.01	&	19.2	&25.3\\
11	&	2007	&	3	&	28	&	1:58:16	&	2454187.58204	&	2240.90&	8.517	&	-24.84	&	19.0	&26.0\\
12	&	2007	&	10	&	3	&	11:45:44	&	2454376.99043&	2430.31&	9.112	&	24.68	&	19.6	&14.2\\
13	&	2007	&	11	&	5	&	8:42:34	&	2454409.86573&	2463.18&	9.216	&	19.94	&	19.1	&18.1\\
14	&	2007	&	12	&	18	&	8:22:37	&	2454452.85368&	2506.17&	9.351	&	4.10	&	18.6	&29.6\\
15	&	2008	&	1	&	10	&	5:30:43	&	2454475.73424	&	2529.05&	9.423	&	-5.74	&	18.7	&31.0\\
16	&	2008	&	2	&	8	&	6:51:57	&	2454504.78953	&	2558.10	&	9.514	&	-17.15	&	19.2	&26.2\\
17	&	2008	&	2	&	29	&	2:18:08	&	2454525.59801		&	2578.91	&	9.580	&	-22.19	&	19.1	&21.3\\
\enddata
\tablenotetext{a}{Based on period = 318.00 days; the phase entries include the orbit number before the decimal point, followed by the phase value. }
\end{deluxetable}


\begin{thebibliography}{}
\bibitem[Allen et al.(1980)]{allen80}Allen, D. A., Hyland, A. R., \& Caswell, J. L. 1980, \mnras, 192, 505
\bibitem[Akashi(2007)]{akashi07}Akashi, M. 2007, preprint, arXiv: 0709.0825
\bibitem[Akashi \& Soker(2008)]{akashi08}Akashi, M., \& Soker, N. 2008, New A, 13, 157
\bibitem[Allard \& Hauschildt(1995)]{allard95}Allard, F., \& Hauschildt, P. H. 1995, \apj, 445, 433
\bibitem[Allard et al.(2001)]{allard01}Allard, F., Hauschildt, P. H., Alexander, D. R., Tamanai, A., \& Schweitzer, A. 2001, \apj, 556, 357
\bibitem[Balick \& Frank(2002)]{balick02}Balick, B., \& Frank, A. 2002, \araa, 90, 439
\bibitem[Baron \& Hauschildt(1998)]{baron98}Baron, E., \& Hauschildt, P. H. 1998, \apj, 495, 370
\bibitem[Baron et al.(1996)]{baron96}Baron, E., Hauschildt, P. H., Nugent, P., \& Branch, D. 1996, \mnras, 283, 297
\bibitem[Bieging et al.(2006)]{bieging06}Bieging, J. H., Schmidt, G. D., Smith, P. S., \& Oppenheimer, B. D. 2006, \apj, 639, 1053
\bibitem[Blocker(1995)]{blocker95}Blocker, T 1995, \aap, 299, 755
\bibitem[Bonacic Marinovic et al.(2008)]{bonacic08}Bonacic Marinovic, A. A., Glebbeek, E., \& Pols, O. R. 2008, \aap, 480, 797
\bibitem[Bujarrabal et al.(2005)]{bujarrabal05}Bujarrabal, V., Castro-Carrizo, A., Alcolea, J., \& Neri, R. 2005, \aap, 441, 1031
\bibitem[]{}Casse, F., \& Keppens, R. 2004, \apj, 601, 90
\bibitem[Cohen et al.(1975)]{cohen75}Cohen, M., Anderson, C. M., Cowley, A. et al. 1975, \apj, 196, 179
\bibitem[Cohen et al.(2004)]{cohen04}Cohen, M., van Winckel, H., Bond, H. E., \& Gull, T. R. 2004, \aj, 127, 2362
\bibitem[De Marco(2006)]{demarco06}De Marco, O. 2006, Proc. IAU Symposium 234, in press, arXiv:0605:626[astro-ph]
\bibitem[de Ruyter et al.(2006)]{deruyter06}de Ruyter, S., van Winckel, H., Maas, T., Lloyd Evans, T., Waters, L. B. F. M., \& Dejonghe, H 2006, \aap, 448, 641
\bibitem[]{}Eisloeffel, J., \& Mundt, R. 1998, \aj, 115, 1554
\bibitem[]{}Ferreira, J., Dougados, C., \& Cabrit, S. 2006, \aap, 453, 785
\bibitem[]{}Hauschildt, P. H. 1992, JQSRT, 47, 433
\bibitem[]{}Hauschildt, P. H. 1993, JQSRT, 50, 301
\bibitem[]{}Hauschildt, P. H., \& Baron, E. 1995, JQSRT, 54, 987
\bibitem[]{}Hauschildt, P. H., Baron, E., Starrfield, S., \& Allard, F. 1996, \apj, 462, 386
\bibitem[]{}Hauschildt, P. H., Baron, E., \& Allard, F. 1997, \apj, 483, 390
\bibitem[]{}Hauschildt, P. H., Lowenthal, D. K., \& Baron, E. 2001, \apjs, 134, 323
\bibitem[]{}Hobbs, L. M., Thorburn, J. A., Oka, T., Barentine, J., Snow, T. P., \& York, D. G. 2004, \apj, 615, 947
\bibitem[]{}Hubeny, I., Agol, E., Blaes, O., \& Krolik, J. H. 2000, \apj, 533, 710
\bibitem[]{}Jura, M., Turner, J., \& Balm, S. P. 1997, \apj, 474, 741
\bibitem[]{}Icke, V. 1981, \apj, 247, 152
\bibitem[]{}Icke, V. 2003, Proc. 209th IAU Symp, Eds. Sun Kwok, Michael Dopita, \& Ralph Southerland, ASP Publ., p. 495 (http://www.strw.leidenuni.nl/~icke)
\bibitem[]{}Kelly, D. M., \& Latter, B. B. 1995, \aj, 109, 1320
\bibitem[]{}Kriz, S., \& Hubeny, I. 1986, Bull. Astron. Inst. Czechosl., 37, 129
\bibitem[]{}Lagadec, E., \& Zijlstra, A. A. 2008, \mnras, 390, L59
\bibitem[]{}Lightman, A. P. 1974, \apj, 194, 419
\bibitem[]{}Livio, M. 1997, Accretion Phenomena and related Outflows, IAU Coll. 163, ASP Conf. Ser. 121, D. T. Wickramasinghe, G. V. Bicknell, \& L. Ferrario, (eds), 845
\bibitem[]{}Livio, M., Salzman, J., \& Shaviv, G. 1979, MNRAS, 188, 1
\bibitem[]{}Mastrodemos, N., \& Morris, M. 1998, \apj, 497, 303
\bibitem[]{}Mastrodemos, N., \& Morris, M. 1999, \apj, 523, 357
\bibitem[]{}Mattsson, L., Wahlin, R., Hoefner, S., \& Eriksson, K. 2008, \aap, 484, L5
\bibitem[]{}Men'shchikov, A. B., Balega, Y. Y., Osterbart, R., \& Weigelt, 1998, New Astron. 3, 601
\bibitem[]{}Men'shchikov, A. B., Schertl, D., Tuthill, P. G., Weigelt, G., \& Yungelson, L. R. 2002, \aap, 393, 867
\bibitem[]{}Morris, M. 1981, \apj, 249, 572
\bibitem[]{}Morris, M. 1987, \pasp, 99, 1115
\bibitem[]{}Nordhaus, J., \& Blackman, E. G. 2006, \mnras, 370, 2004
\bibitem[]{}Osterbart, R., Langer, N., \& Weigelt, G. 1997, \aap, 325, 609
\bibitem[]{}Papaloizou, J., \& Pringle, J. E. 1977, MNRAS, 181, 441
\bibitem[]{}Price. S. D., \& Walker, R. G.  1976, The AFGL Four Color Infrared Sky Survey: Catalog of Observations at 4.2, 11.0, 19.8, and 27.4 Micrometers (AFGL-TR-0208) (Hanscom , MA: Opt. Phys. Div.)
\bibitem[]{}Pringle, J. E. 1981, ARAA, 19, 137
\bibitem[]{}Pudritz, R. E., \& Norman, C. A. 1983, \apj, 274, 677
\bibitem[]{}Ramstedt, S., Schoeier, F. L., Olofsson, H., \& Lundgren, A. A. 2008, \aap, 487, 645
\bibitem[]{}Regos, E., Bailey, V. C., \& Mardling, R. 2005, MNRAS, 358, 544
\bibitem[]{}Rupen, M. P.,  Mioduszewski, A. J. , \& Sokoloski, J. L. 2008, \apj, 688, 559
\bibitem[]{}Sahai, R., Findeisen, K., Gil de Paz, A. \& Sanchez Contreras, C. 2008, \apj, in press, arXiv0807:1944v1 [astro-ph]
\bibitem[]{}Schmidt, G. D., Cohen, M., \& Margon, B. 1980, \apj, 239, L133
\bibitem[]{}Schmidt, G. D., \& Witt, A. N. 1991, \apj, 383, 698
\bibitem[]{}Sepinsky, J. F., Willems, B., \& Kalogera, V. 2007a, \apj, 660, 1624
\bibitem[]{}Sepinsky, J. F., Willems, B., Kalogera, V., \& Rasio, F. A. 2007b, \apj, 667, 1170
\bibitem[]{}Siodmiak, N., Meixner, M., Ueta, T., Sugerman, B. E. K., Van de Steene, G. C., \& Szczerba, R. 2008, \apj, 677, 382
\bibitem[]{}Sitko, M. L., 1983, \apj, 265, 848
\bibitem[]{}Sitko, M. L., Bernstein, L. S., \& Glinski, R. J. 2008, \apj, 680, 1426
\bibitem[]{}Soker, N. 2000, \aap, 357, 557
\bibitem[]{}Soker, N. 2005, \aj, 129, 947
\bibitem[]{}Soker, N. 2008, New A, 13, 491
\bibitem[]{}Soker, N., \& Livio, M. 1994, \apj, 421, 219
\bibitem[]{}Szczerba, R., Siodmiak, N., Stasinska, G., \& Borkowski, J. 2007, \aap, 469, 799
\bibitem[]{}Thorburn, J. A., et al. 2003, \apj, 584, 339
\bibitem[]{}Tuthill, P. G., Men'shchikov, A. B., Schertl, D., Monnier, J. D., Danchi, W. C., \& Weigelt, G. 2002, \aap, 389, 889
\bibitem[]{}van Winckel, H. 2003, \araa, 41, 391, 868
\bibitem[]{}van Winckel, H. 2007, Baltic Astron., 16, 112
\bibitem[]{}van Winckel, H., Lloyd Evans, T, Reyniers, M., Deroo, P., \& Gielen, C. 2006, Mem. S. A. It. , 77, 943
\bibitem[]{}van Winckel, H., Waelkens, C., \& Waters, L. B. F. M. 1995, \aap, 293, L23
\bibitem[]{}Vijh, U. P., Witt, A. N., York, D. G., Dwarkadas, V., Woodgate, B. E., \& Palunas, P. 2006, \apj, 653, 1336
\bibitem[]{}Waelkens, C., van Winckel, Waters, L. B. F. M., \& Bakker, E. J. 1996, \aap, 314, L17
\bibitem[]{}Wang, S. et al. 2003, Proc. SPIE, 4841, 1145
\bibitem[]{}Waters, L. B. F. M. et al. 1998, Nature, 391
\bibitem[]{}Witt, A. N., \& Boroson, T. A. 1990, \apj, 355, 182
\bibitem[]{}Witt, A. N., Gordon, K. D., Vijh, U. P., Sell, P. H., Smith, T. L., \& Xie, R.-H. 2006, \apj, 636, 303
Woitke, P. 2006, A\&A, 460, L9
\bibitem[]{}Ymamura, I., Dominik, C., de Jong, T., Waters, L. B. F. M., \& Molster, F. J. 2000, \aap, 363, 629
\end{thebibliography}
\end{document}